\documentclass[12pt]{article}
\usepackage{jheppub}
\usepackage{amssymb,amsfonts}

\usepackage{wrapfig}
\usepackage{epsfig}
\usepackage[T1]{fontenc}
\usepackage[utf8]{inputenc}
\usepackage{lmodern}
\usepackage{float}
\usepackage{placeins}
\usepackage{dsfont}
\usepackage{arydshln}

\usepackage[dvipsnames]{xcolor}
\usepackage{tikz}

\def\be{\begin{eqnarray}}
\def\ee{\end{eqnarray}}
\newcommand{\nn}{\nonumber}
\newcommand\para{\paragraph{}}
\newcommand{\ft}[2]{{\textstyle\frac{#1}{#2}}}
\newcommand{\eqn}[1]{(\ref{#1})}

\newcommand\bgamma{\mbox{\boldmath $\gamma$}}

\newcommand\bsigma{\mbox{\boldmath $\sigma$}}

\newcommand{\bmu}{{\boldsymbol{\mu}}}
\newcommand{\brho}{{\boldsymbol{\rho}}}
\newcommand{\bSigma}{{\boldsymbol{\Sigma}}}

\newcommand{\ra}{\rightarrow}

\newcommand\br{{\bf r}}
\newcommand{\varep}{\varepsilon}

\def\Dslash{\,\,{\raise.15ex\hbox{/}\mkern-12mu D}}
\def\Dbarslash{\,\,{\raise.15ex\hbox{/}\mkern-12mu {\bar D}}}
\def\delslash{\,\,{\raise.15ex\hbox{/}\mkern-9mu \partial}}
\def\delbarslash{\,\,{\raise.15ex\hbox{/}\mkern-9mu {\bar\partial}}}
\def\pslash{\,\,{\raise.15ex\hbox{/}\mkern-9mu p}}
\def\calDslash{\,\,{\raise.15ex\hbox{/}\mkern-12mu {\cal D}}}

\newcommand{\Z}{{\mathds Z}}

\newcommand{\R}{{\mathds R}}

\newcommand{\bk}{{\bf k}}

\newcommand{\bs}{{\bf s}}
\newcommand{\bt}{{\bf t}}

\newcommand{\horridt}{\mathsf{T}}

\newcommand{\whQ}{\widehat{Q}}
\newcommand{\whA}{\widehat{A}}
\newcommand{\whB}{\widehat{B}}

\def\lae{\mathrel{\mathop{\smash{\lower .5 ex \hbox{$\stackrel<\sim$}}}}}
\def\lae{\mathrel{\mathop{\smash{\lower .5 ex \hbox{$\stackrel>\sim$}}}}}

\usepackage[title]{appendix}

\title{Onsager Algebras and Lattice Fermions}

\author[a]{Ioanna Kourkoulou}
\author[a,b]{and David Tong}

\affiliation[a]{Department of Applied Mathematics and Theoretical Physics \\ University of Cambridge, CB3 0WA, UK}
\affiliation[b]{CERN, Theoretical Physics Department, Geneva, Switzerland}

\emailAdd{ik344@cam.ac.uk}
\emailAdd{d.tong@damtp.cam.ac.uk}

\abstract{We study the exact Onsager algebra of free lattice fermions in staggered, Wilson, and overlap incarnations, working  in  both Hamiltonian and Euclidean formulations in four spacetime dimensions. In each case, we show that the exact lattice algebra emanates the full chiral symmetry of the low-energy theory: if there are $N$ massless Dirac fermions in the continuum, the Onsager algebra reduces to a $U(2N)$ chiral symmetry. The details of which generators are quantised or ultra-local differ between formulations. There is also a trade-off between the locality of generators and the maximum compact subgroup of the Onsager algebra.}

\begin{document}
\pagestyle{plain} \setcounter{page}{1}
\newcounter{bean}
\baselineskip16pt \setcounter{section}{0}

\maketitle


\section{Introduction}

Putting fermions on the lattice requires compromise. Wilson fermions eliminate doublers at the expense of breaking the naive axial symmetry. Staggered fermions retain some doublers -- usually called {\it tastes} -- while preserving a small  remnant of the chiral symmetry. Overlap fermions eliminate the doublers and preserve a chiral symmetry, but its conserved charge is momentum-dependent and unquantised. Moreover, the Hamiltonian has a weakened notion of locality (at least in higher dimensions). Regardless of the approach, realising chiral symmetry on the lattice involves an interplay between removing doublers, quantisation, and locality. You can't have them all; you need to choose. 

\para
These obstacles are enshrined in the Nielsen-Ninomiya theorem \cite{nn1,nn2,friedan} and its generalisations \cite{xf,anton}. 
The hope is that, among these various compromises, there is a route --  involving suitable interactions -- to evading these obstacles and finding a practical way of simulating chiral gauge theories, like the Standard Model.  (Overlap fermions provide an important but, so far, impractical way of simulating chiral theories on the lattice \cite{l1,craig}.) For classic reviews, see \cite{luscherreview,kaplan}.

\para
Recently, it has become apparent that some fermion lattice models admit a larger symmetry than previously realised. In many situations, this symmetry is an infinite-dimensional {\it Onsager algebra}, so called because it is a variant of the algebra Onsager encountered in his solution to the 2d Ising model. A particularly simple example occurs for the $d=1+1$ massless staggered fermion \cite{shuheng}. In the infrared, the generators of this algebra reduce to the familiar $U(1)_V\times U(1)_A$ symmetries of a massless Dirac fermion, but they manifest themselves on the lattice in a novel manner: both the vector charge $Q_V$ and the axial charge $Q_A$ are quantised, but they do not commute with each other -- a result first noted in \cite{thack1,thack2} --  reflecting their mixed 't Hooft anomaly. 

\para
In the language of \cite{nati1,nati2}, the continuum chiral symmetry is  {\it emanant}, rather than emergent, meaning it is already lurking in the microscopic algebra, but takes a different form after flowing to the infrared. 

\para
Variations on  the Onsager algebra, with different emanant symmetries, have been found in many other lattice models, including chiral Potts models \cite{perk}, XX spin chains \cite{fendley,minami,tdual}, Weyl semi-metals \cite{ryan}, graphene \cite{valley} and staggered fermions \cite{fermi,simon,aoki,onogi,misumi} in various dimensions. See \cite{xu,yam,elijah} for related work.

\para
The purpose of this paper is to look at the Onsager algebra for staggered, Wilson and overlap fermions. We study these in both Hamiltonian  (i.e. $d=3+1$ dimensions) and Euclidean  ($d=4+0$) incarnations. In each case, we ask: how much of the chiral symmetry is emanant? In each case, the answer is: all of it.

\para
The details differ for each kind of fermion. Here we summarise the key points.

\para
The Onsager algebra for staggered fermions in the $d=3+1$ Hamiltonian formulation was previously studied in \cite{simon,aoki,onogi,misumi}. At low energies, a single staggered fermion gives rise to two Dirac fermions. When the mass vanishes, it was shown that the Onsager algebra emanates a $U(2)_L \times U(2)_R$ chiral symmetry. We revisit this result in  Section \ref{3dsec} and show that there is a larger symmetry: after conjugating the right-handed fermions, all four Weyl fermions can be viewed as having the same chirality, and the full emanant symmetry is $U(4)$. Moreover, with $N_f$ staggered fermions, the emanant symmetry is $U(4N_f)$. 

\para
The Onsager algebra is infinite-dimensional. It's natural to ask what the maximal compact subalgebra is. If we restrict to ultra-local operators then, with $N_f$ staggered fermions (meaning $2N_f$ Dirac fermions in the infrared), the answer is the obvious $U(N_f)$ symmetry. But if we relax the locality condition, and instead allow generators that are merely exponentially local, rather than ultra-local, we find a larger $U(2N_f)$ symmetry acting on the lattice fermions. 

\para
In Section \ref{w3dsec}, we turn to Wilson fermions. Here there is an Onsager algebra that exists for any mass, but in the gapless case it has a chiral extension. 
For $N_f$ Wilson fermions, this Onsager algebra again collapses to the full chiral symmetry, which is now $U(2N_f)$. This time, however, a compromise must be made. There is a choice of certain chiral generators in the ultraviolet. One choice is ultra-local, but unquantised; the other quantised but only exponentially local. In the latter case, we can find an exact $Sp(N_f)$ lattice symmetry.

\para
In Section \ref{over3dsec}, we discuss overlap fermions. The story is almost identical to Wilson fermions, with an extended Onsager algebra that, for $N_f$ fermions,  emanates the full $U(2N_f)$ symmetry. Again, one can construct a compact $Sp(N_f)$ subgroup in the ultraviolet.

\para
The first few sections are devoted to Hamiltonian formulations. 
In Section \ref{4dsec}, we turn to fermions on Euclidean lattices. We focus on staggered fermions. Now, one staggered fermion describes four tastes. There is again an extension of the Onsager algebra which, in the infrared, collapses to $U(8)$. Wilson and overlap fermions also admit an Onsager algebra that emanates the full chiral symmetry, but the detailed calculations bring nothing particularly new to the table and are omitted.
We summarise all these results in Section \ref{summarysec}.

\para
These results hold only for free fermions and their significance for interacting theories remains unclear. Very little of the Onsager algebra survives the standard gauging procedure. With just a single fermion, this can be viewed as a virtue: because the precursors of $Q_V$ and $Q_A$ do not commute on the lattice, gauging the $U(1)_V$ vector symmetry necessarily breaks the axial charge, providing a lattice realisation of the ABJ anomaly \cite{shuheng,aoki}. (A complementary perspective on discrete chiral symmetry in the Schwinger model can be found in \cite{dempsey}.) 

\para
However, the extension to $N_f$ fermions adds a wrinkle to this story. For $N_f$ free staggered fermions in $d=1+1$, the emanant symmetry is $U(N_f)_L \times U(N_f)_R$. After gauging the vector $U(1)_V$, this is reduced to $SU(N_f)_V$; the full chiral symmetry is lost microscopically, but the anomaly is responsible for only the loss of $U(1)_A$. The enhancement of the microscopic $SU(N_f)_V$ to the chiral $SU(N_f)_L\times SU(N_f)_R$ in the continuum is no longer emanant, merely emergent. Similar statements hold in higher dimensions.

\para
Nonetheless, our hope is that identifying exact chiral symmetries  of free fermions on the lattice may provide a useful starting point for  future constructions. 

\para
We include two appendices. In Appendix \ref{2dsec}, we revisit staggered fermions in $d=2+1$, following \cite{fermi}: we show that there is an emanant $SO(4)$ symmetry, which rotates the four massless Majorana fermions. In Appendix \ref{gwosec}, we review an old construction of Mandula \cite{mandula0,mandula}, who showed that there is a different Onsager algebra associated with the Ginsparg-Wilson relation, albeit one that does not commute with the Hamiltonian.

\section{Hamiltonian Staggered Fermions}\label{3dsec}

We start with staggered fermions in $d=3+1$ dimensions, also known in the condensed matter literature as $\pi$-flux models. As we review, at low energies, the theory gives rise to two gapless Dirac (i.e. four-component) fermions, but with different components sitting at different corners of the Brillouin zone. Equivalently, these can be viewed as four Weyl fermions.

\para
The Onsager algebra in this model was studied previously in \cite{simon,aoki,onogi,misumi}. (Strictly, it should be referred to as a magnetic Onsager, reflecting a proliferation of annoying minus signs that appear in the algebra \cite{fermi}). The algebra elements are exact lattice symmetries. In the infrared, it was shown that these symmetries emanate a $U(2)_L \times U(2)_R$ chiral symmetry. Here we  look more closely and show that, in fact, the infrared symmetry is somewhat larger and the Onsager algebra emanates a $U(4)$ symmetry, which rotates four Weyl fermions. The overall $U(1)\subset U(4)$ is identified as the axial $U(1)_A$ and, as pointed out in \cite{aoki}, is unquantised, a fact that follows on general grounds \cite{xf}. Two of the three  other Cartan elements of $U(4)$ are quantised.

\para
We take a  cubic lattice with sites labelled by $\br = (x,y,z)\in \Z^3$. The physical lattice spacing is $a$.  There is a single-component fermion $c_{\br}$  (i.e. a single qubit) on each site, with anti-commutation relations
\be \{c_{\br} ,c_{\br'}\} = \{c^\dagger_\br,c^\dagger_{\br'}\}=0 \ \ \ {\rm and}\ \ \ \{c_\br,c^\dagger_{\br'}\} = \delta_{\br,\br'}\ \label{ccom}\ .\ee
We will consider an infinite lattice, but the construction also holds if we take the lattice to have size $L_x,L_y,L_z\in 2\Z$ and impose periodic boundary conditions on the fermions. 

\para
The Hamiltonian is
\be H = \frac{i}{2}\sum_\br \sum_{i=1}^3\eta_i(\br)c_\br^\dagger\left(c_{\br + \hat{\bf i}} - c_{\br - \hat{\bf i}}\right)\label{3dham}\ee
where $\hat{\bf i}$ denotes a shift by a single lattice site in the $i=1,2,3$ direction and $\eta_i(\br)$ are a bunch of phases which implement the staggering along a link $\br = (x,y,z)$,
\be \eta_1(\br) = -1\ ,\ \ \ \eta_2(\br) = (-1)^x\ ,\ \ \ \eta_3(\br) = (-1)^{x+y}\ .\ee
\noindent
\begin{wrapfigure}{r}{0.2\textwidth}
\centering
\includegraphics[width=.7\linewidth]{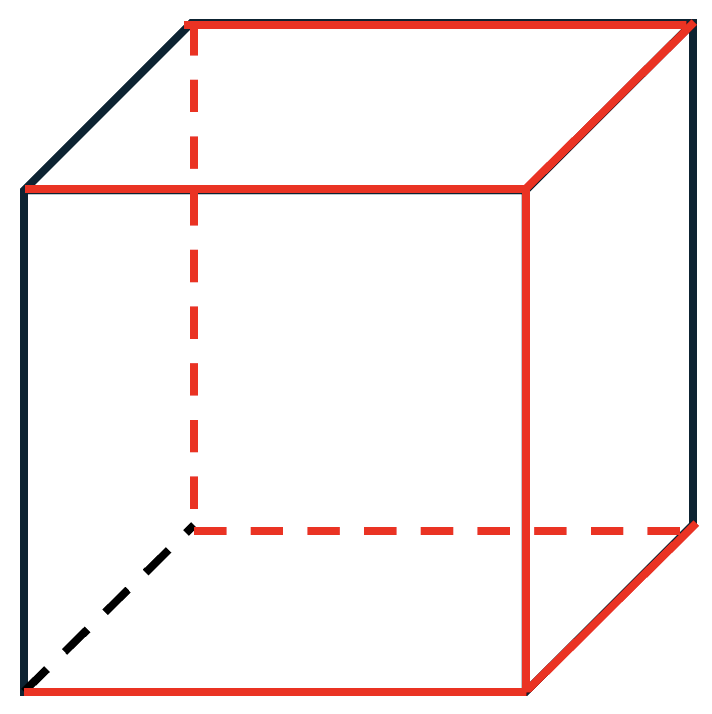}
\end{wrapfigure}
These are designed so that traversing the links of any plaquette of a unit cube gives rise to an overall minus sign. In other words, we can think of this as a $\pi$-unit of flux through each plaquette. An example of such a unit cube is shown on the right; links with plus signs are in black, and those with minus signs in red.

\para
The staggered minus signs mean that the  Hamiltonian isn't invariant under the naive translations $\br \ra \br + \hat{\bf i}$. It is, however, straightforward to patch this up by introducing compensating minus signs into the translation operators. It's simple to check that the Hamiltonian is invariant under the staggered translations
\be T_ic_\br T_i^{-1} = \xi_i(\br) c_{\br + \hat{\bf i}}\ee
where $\xi_i(\br)$ are a collection of signs defined by
\be \xi_1(\br) = (-1)^{y+z}\ ,\ \ \ \xi_2(\br) = (-1)^z\ ,\ \ \ \xi_3(\br) = 1\ .\label{3dxi}\ee
This tells us that the true unit cell of the problem comes from taking two lattice sites in the $x$- and $y$-directions, while leaving the $z$-direction unchanged. While that's true, the physics is more transparent if we treat all directions on an equal footing and take the unit cell to be a cube (rather than a point). To this end, we write
\be {\bf r} = 2{\bf n} + \brho\ .\ee
Here ${\bf n}$ picks the left-hand corner of a cube while $\brho$ has components  $\rho_i=0,1$ and ranges over the eight different vertices. We then collect together the fermion operators attached to each vertex to form an 8-component spinor. This will be the object we work with in what follows.

\para
It's simplest to first work in momentum space. Because we've doubled the unit cell in each direction, the Brillouin zone is halved, with momenta lying in the range
\be -\frac{\pi}{2a} < k_i \leq \frac{\pi}{2a}\ \ \ i=1,2,3\ .\label{bz3d}\ee
We then write the 8-component spinor in momentum space as
%
%
\be \psi(\bk) = \big( c_{000}(\bk) ,  c_{001}(\bk) ,  c_{010}(\bk) ,   c_{011}(\bk) ,  c_{100}(\bk) ,  c_{101}(\bk) , c_{110}(\bk) ,  c_{111}(\bk) \big)^T \ .\label{psi3d}\ee
Here the components are Fourier transforms of only those fermions that sit on one corner of the cube
\be c_{\brho}(\bk) = \sqrt{\frac{8}{N}} \sum_{\bf n} e^{-i\bk\cdot (2{\bf n}+\brho) a}c_{2{\bf n}+\brho}\ .\ee
The choice of ordering in \eqn{psi3d} is deliberate. We will find collections of $8\times 8$ Hermitian matrices acting on these fermions and spend some time disentangling their meaning. These matrices can be built from tensor products of Pauli matrices which, following the notation used in quantum information, we denote as  $X$, $Y$, and $Z$. We  append a subscript to denote where in the tensor product the matrix lives. So, for example,
\be X_1 = X \otimes \mathds{1}\otimes \mathds{1}\ ,\ \ \ X_2 = \mathds{1}\otimes X \otimes \mathds{1}\ ,\ \ \ Y_1Z_2 = Y\otimes Z \otimes \mathds{1}\ .\ee
The order  in \eqn{psi3d} is such that the $i^{\rm th}$ entry in the product acts on the $i^{\rm th}$ index of the subscript $\brho$. 

\para
There is an important implication of working in the reduced Brillouin zone \eqn{bz3d}: the 8-component spinor $\psi(\bk)$ is not periodic in $\bk$. To see this, note that
\be c_{\brho}\left(\bk + \frac{\pi}{a} \hat{\bf i}\right) = \sqrt{\frac{8}{N}}\sum_{\bf n} e^{-i(\bk + (\pi/a)\hat{\bf i})\cdot(2{\bf n} + \brho)a} c_{2{\bf n}+\brho} = (-1)^{\rho_i} c_\brho(\bk)\ .\ee
That additional $(-1)^{\rho_i}$ means that those $c_\brho$ with $\rho_i=0$ are periodic, while those with $\rho_i=1$ are anti-periodic. That's very easy to implement with our notation above: we  must have
\be \psi\left(\bk  + \frac{\pi}{a}\hat{\bf i}\right) = Z_i \,\psi(\bk)\ .\label{3dperiod}\ee
The Hamiltonian \eqn{3dham} takes a simple form when written in momentum space in terms of $\psi(\bk)$. First note that the Hamiltonian couples $c_{\br}^\dagger$ to $c_{\br \pm \brho}$ for some $\brho$. When written in terms of the 8-component spinor $\psi$, this turns into a coupling $\psi^\dagger X_i \psi$. In addition, the staggered signs $\eta_i(\br)$ introduce a  $Z_1$ matrix for hopping in the $y$-direction, and a $Z_1Z_2$ matrix for hopping in the $z$-direction. The upshot is that the Hamiltonian in momentum space becomes
\be H = -\int \frac{d^3k}{(2\pi)^3}\ \psi^\dagger (\bk)\,h(\bk)\,\psi(\bk)\ee
where the integral is over the reduced Brillouin zone \eqn{bz3d} and 
\be h(\bk) = -\sin (k_1a)\,X_1 + \sin (k_2a) \, Z_1 X_2 + \sin (k_3a)\,Z_1Z_2X_3\ .\label{hk3d}\ee
The matrices in this Hamiltonian  anti-commute with each other and square to one, allowing us to immediately read off the spectrum
\be E_\pm(\bk) = \pm \sqrt{\sum_{i=1}^3 \sin^2 (k_ia)}\ .\ee
There is just a single gapless point in the reduced Brillouin zone, at $\bk = 0$. This can be viewed as two gapless Dirac fermions.  In the original Brillouin zone, the different components of these fermions come from different points  where $k_i=0$ or $\pi/a$. 

\para
These Dirac fermions can also be viewed as four gapless Weyl (i.e. 2-component) fermions, two left-handed and two right-handed. To see how this comes about, first note that a massless Weyl fermion has Hamiltonian $h(\bk) = \pm \bsigma\cdot\bk$ where the $+$ sign is for right-handed and the $-$ sign for left-handed. Close to the gapless point $\bk=0$, we can write \eqn{hk3d} in a similar form, with
\be h(\bk) = a\Gamma\,\bSigma \cdot \bk \ ,\label{3dlow}\ee
where 
\be \Gamma = X_1 Y_2 X_3\ ,\ \ \ \Sigma^1 = -Y_2X_3\ ,\  \ \ \Sigma^2 = -Y_1Z_2X_3\ ,\ \ \ \Sigma^3 = Y_1X_2\ .\label{3dspin}\ee
Here $\Gamma$ plays the role of an $8\times 8$ $\gamma^5$ matrix. It obeys $\Gamma^2 = \mathds{1}$  and has four $+1$ eigenvalues, corresponding to left-handed Weyl spinors, and four $-1$ eigenvalues, corresponding to right-handed spinors. Meanwhile, the matrices $\Sigma^i$ obey the same algebra as Pauli matrices
\be \Sigma^i\Sigma^j = \delta^{ij} + i\epsilon^{ijk}\Sigma^k\ \ee
and $[\Gamma,\Sigma^i]=0$. These are the spinor matrices for the staggered fermion. 

\para
The low-energy Hamiltonian \eqn{3dlow} is invariant under the full $U(2)_L \times U(2)_R$ chiral symmetry of two Dirac (or four Weyl) massless fermions. The $U(1)_V$ is simply the overall phase rotation $\psi \ra e^{i\alpha}\psi$, while the axial $U(1)_A$ is generated by $\Gamma$. In addition, it's simple to check that the following matrices
\be \tau^1 = X_2Y_3\ ,\ \ \ \tau^2= -X_1Z_2Y_3\ ,\ \ \ \tau^3=X_1Y_2\label{taste}\ee
obey $[\tau^i,\tau^j] = 2i\epsilon^{ijk} \tau^k$ and $[\tau^i,\Sigma^j]=0$. The matrices $\tau^i$ act in the obvious way on the 8-component spinor, $\psi \ra \tau^i \psi$, and generate the $SU(2)_V$ group of transformations.  These are {\it taste} generators in the language of staggered fermions. The two commuting $SU(2)_L\times SU(2)_R$ symmetries are generated by $\ft12 (1\pm \Gamma)\tau^i$.

\para
At first glance, it appears that all these are also  symmetries of the full lattice Hamiltonian \eqn{hk3d}; after all,  the $\tau^i$ commute with each of the matrices in $h({\bf{k}})$. But that's not the only requirement: the transformations must be consistent with the periodicity condition \eqn{3dperiod} and, because  $[\tau^i,Z^j]\neq 0$, none of the would-be $SU(2)$ symmetries are permissible when acting on the full Brillouin zone.

\para
There are different ways forward. First,  it's possible to dress the symmetry generators with momentum dependence, to ensure that they have the correct periodicity. For example, we can define $\hat{\tau}^1 = \cos(k_2a) \cos(k_3a)\tau^1$, $\hat{\tau}^2 = \cos(k_1a)\cos(k_3a) \tau^2$ and $\hat{\tau}^3 = \cos(k_1a)\,\cos(k_2a)\tau^3$, each of which is a symmetry of the Hamiltonian and respects the periodicity condition. The price to pay is that the charges are no longer integer quantised, and they no longer close into the $su(2)$ algebra. 

\para
A second alternative is to take the generators to be constant, but only piecewise continuous (which is a cheap way of saying discontinuous)  across the Brillouin zone. If we define $q_j(\bk) = {\rm sign}[\cos(k_ja)]$, then we could take $\hat{\tau}^1 = q_2q_3\tau^1$, $\hat{\tau}^2 = q_1q_3\tau^2$  and $\hat{\tau}^3 = q_1q_2\tau^3$. This now has the correct  behaviour in momentum space and closes to an $su(2)$ algebra with  quantised charges. The price we pay, however, is that the transformation is non-local in position space. This kind of transformation has been described  in the context of lattice QCD -- see, for example, \cite{rothe}.  

\para
The Onsager algebra allows us to have our cake and eat it (at least for the free fermions discussed here). The staggered fermion lattice Hamiltonian admits a  magnetic Onsager algebra  \cite{simon,aoki,onogi,misumi,fermi}.  Here we rederive this algebra and show that, viewed appropriately, it actually captures the full $U(4)$ chiral symmetry in the infrared.

\subsection{The Onsager Algebra}

The Onsager algebra is best seen by working with Majorana fermions \cite{shuheng}. We write
\be c_\br = \frac{1}{2}(a_\br + i b_\br)\ee
with $a_\br^\dagger = a_\br$ and $b_\br^\dagger =b_\br$. These Majorana components satisfy
\be \{a_\br,a_{\br'}\} = \{b_\br,b_{\br'}\} = 2\delta_{\br,\br'} \ \ \ {\rm and}\ \ \ \{a_\br,b_{\br'}\}=0\ .\ee
Crucially, the Majoranas $a_\br$ and $b_\br$ decouple in the Hamiltonian \eqn{3dham}, which becomes
\be H = \frac{i}{4}\sum_\br \sum_{i=1}^3\eta_i(\br)\left(a_\br a_{\br +\hat{\bf i}} + b_\br b_{\br + \hat{\bf i}}\right)\ .\label{3dabham}\ee
This is the property that ensures the existence of an Onsager algebra. In particular, the theory is invariant under independent translations of $a_\br$ and $b_\br$. For example, we can introduce the translation operator $\horridt_i$ which leaves  $a_\br$ untouched, so $\horridt_ia_\br \horridt_i^{-1} = a_\br$  and acts only on $b_\br$ as
\be \horridt_i b_\br \horridt_i^{-1} = \xi_i(\br) b_{\br + \hat{\bf i}} \ee
where $\xi_i(\br)$ are the collection of minus signs defined in \eqn{3dxi}. For a finite translation by $\bs=(s_1,s_2,s_3)$, we define 
\be \horridt_\bs  = \horridt_3^{s_3} \horridt_2^{s_2} \horridt_1^{s_1} \ \ \ \Longrightarrow\ \ \ \horridt_\bs b_\br \horridt_\bs^{-1} = \xi_\bs(\br)b_{\br + \bs}\ \ {\rm with}\ \ \xi_\bs(\br) = (-1)^{s_1(y+z) + s_2z} \ .\ \ \ \ \label{3dhorrid}\ee
We can use this to construct an infinite number of conserved quantities. We start with the manifest particle number symmetry, arising from $c_\br \ra e^{-i\alpha}c_\br$. The associated charge is 
\be Q_{\bf 0} = \sum_{\br} \left(c_\br^\dagger c_\br - \frac{1}{2}\right) = \frac{i}{2}\sum_\br a_\br b_\br\ .\label{3dq0}
\ee
Importantly, this is quantised: $Q_{\bf 0} \in \Z$. The $-\ft12$ in the definition ensures that the charge is measured relative to half-filling. But with one charge in hand, we can generate an infinite collection of further charges by acting with the translation operators $\horridt_\bs$. The result is
\be Q_\bs = \horridt_\bs Q_{\bf 0} \horridt_\bs^{-1} = \frac{i}{2} \sum_\br \xi_\bs(\br) a_\br b_{\br+\bs}\ \ \ {\rm with}\ \ \ \xi_\bs(\br) = (-1)^{s_1(y+z) + s_2z}\ .\label{3dq}\ee
Each of these obeys $[H,Q_\bs]=0$ and each is quantised with spectrum $Q_\bs \in \Z$ or, equivalently, $e^{2\pi i Q_\bs} = 1$. (When working on a  lattice of size $L_x, L_y, L_z\in 2\Z$, with periodic boundary conditions, we have $Q_{\bs + (L_x,0,0)} = Q_{\bs + (0,L_y,0)} = Q_{\bs + (0,0,L_z)} = Q_\bs$.)

\para
The operators $Q_\bs$ do not commute. The full algebra includes the additional operators
\be A_\bs &=& \frac{i}{2}\sum_\br \left[ \xi_\bs(\br) a_\br a_{\br +\bs} - \delta_{\bs,{\bf 0}}\right] \nn\\
B_\bs &=& \frac{i}{2}\sum_\br \left[ \xi_\bs(\br) b_\br b_{\br +\bs} - \delta_{\bs,{\bf 0}}\right]  \ .\label{3dAB}\ee
These too commute with the Hamiltonian but, in contrast to the $Q_\bs$, the spectrum of operators $A_\bs$ and $B_\bs$ is {\it not} quantised. 
Commutators of these generators close, but only with a flurry of annoying
minus signs. These are best incorporated by introducing
\be \epsilon(\bs,\bt) = (-1)^{s_2t_1 + s_3(t_1+t_2)}\ .\ee
This has the property that it combines with the minus signs in the generators, as  $\xi_\bs(\br)\xi_\bt(\br+\bs)=\epsilon(\bs,\bt)\xi_{\bs+\bt}(\br)$.
The commutation relations are then 
\be [Q_\bs,Q_\bt] &=&-i\epsilon(\bt,\bt) \left[\epsilon(\bs,\bt) A_{\bs-\bt} -\epsilon(\bt,\bs)B_{\bs-\bt}\right] \nn\\
{[}Q_\bs,A_\bt{]} &=&i\epsilon(\bt,\bs) \left[\epsilon(\bt,\bt)Q_{\bs-\bt}-Q_{\bs+\bt}\right]\nn\\
{[}Q_\bs,B_\bt{]} &=&-i\epsilon(\bs,\bt) \left[\epsilon(\bt,\bt)Q_{\bs-\bt}-Q_{\bs+\bt}\right]\ \label{3dons}\\
{[}A_\bs,A_\bt{]} &=&i\left[\epsilon(\bs,\bt)-\epsilon(\bt,\bs)\right] \left[A_{\bs+\bt}-\epsilon(\bt,\bt)A_{\bs-\bt}\right] \nn\\
{[}B_\bs,B_\bt{]} &=&i\left[\epsilon(\bs,\bt)-\epsilon(\bt,\bs)\right] \left[B_{\bs+\bt}-\epsilon(\bt,\bt)B_{\bs-\bt}\right]\nn\ee
together with  $[A_\bs,B_\bt]=0$.  This algebra was previously derived in  \cite{aoki}, albeit in different notation.  It was referred to as the {\it magnetic Onsager algebra} in \cite{fermi} because it differs from the usual Onsager algebra by that flurry of minus signs. Strictly, it is a magnetic ${\rm Ons}_3$ algebra\footnote{A better name might be the MINUS Onsager algebra, with MINUS standing for {\it many irritating negative unwanted signs}.}, with the subscript reflecting the fact that the labels $\bs,\bt \in \Z^3$.

\subsubsection*{The Infrared Limit}

In momentum space, we write the 8-component spinor as  $\psi(\bk) = \ft12 (\alpha(\bk) + i\beta(\bk))$ with  $\alpha(\bk)^\dagger = \alpha(-\bk)^T$ and $\beta(\bk)^\dagger = \beta(-\bk)^T$. Then the charges  \eqn{3dq} become
\be Q_\bs =  \frac{i}{2} \int \frac{d^3k}{(2\pi)^3} \ \alpha(\bk)^\dagger e^{ia\bk\cdot \bs} (X_1Z_2Z_3)^{s_1}(X_2Z_3)^{s_2}X_3^{s_3} \beta(\bk)\ .\label{3dqmom}\ee
Now we're in a good position to take the low-energy limit. Shifts by a fixed, finite lattice spacing are unimportant in the infrared, except insomuch as they affect the spinor structure. The phase $e^{ia\bk\cdot\bs}$ becomes unimportant in the infrared, and the only remnant of the lattice spacing lies in the matrix structure of the charges. Because all these matrices square to one, only the parity of $s_i$ is important, leaving a collection of eight charges, each of which takes the form
\be Q_\bs \ra \frac{i}{2} \int d^3x\ \alpha^T \Theta_\bs \beta \ee
Each of the $8\times 8$ matrices $\Theta_\bs$ can be written in terms of the $SU(2)$ taste matrices $\tau^i$, defined in \eqn{taste}, and the $\gamma^5$ proxy $\Gamma$ defined in \eqn{3dspin}: they are
\be \Theta_{(0,0,0)} = \mathds{1} \ , \ \  \ \Theta_{(1,0,0)} = \Gamma\tau^1\ ,\ \  \ \Theta_{(0,1,0)} = \Gamma \tau^2\ ,\ \ \  \Theta_{(0,0,1)} = \Gamma \tau^3
\nn\\
\Theta_{(0,1,1)} = i\tau^1 \ , \ \ \ \Theta_{(1,0,1)} =-i\tau^2\ ,\ \  \ \Theta_{(1,1,0)} = i\tau^3\ ,\ \ \  \Theta_{(1,1,1)} = i\Gamma \ .\label{3dtheta}\ee
The matrices on the first line are all symmetric; those in the  second line are anti-symmetric. Only the latter survive when we take the infrared limit  in the  $A_\bs$ and $B_\bs$, which are
\be A_\bs \ra  \frac{i}{2}\int d^3x\ \alpha^T \Theta_\bs \alpha\ \ \ {\rm and} \ \ \ B_\bs  \ra  \frac{i}{2}\int d^3x\ \beta^T \Theta_\bs \beta \ .\label{3dabir}\ee
This means that we have 16 generators in total: eight from $Q_\bs$ and a further four each from $A_\bs$ and $B_\bs$ (which appear in the algebra in the combinations of  $A_\bs \pm B_\bs)$. It's tempting to think that these are the 16 generators of $U(4)$. We now show that this is indeed the case.

\para
First we look at the $Q_\bs$ in the first line of \eqn{3dtheta}. Because these are symmetric matrices, when we write them in terms of complex fermions $\psi= \ft 12(\alpha + i\beta)$, the $\alpha^T\cdot \alpha$ and $\beta^T\cdot \beta$ cross-terms vanish and they all preserve particle number. They are
\be Q_{\bf 0} = \int d^3x\ :\psi^\dagger\psi:\ \ \ {\rm and}\ \ \ Q_i = \int d^3x\ \psi^\dagger\Gamma \tau^i\psi \ .\ee
Here the colons in $Q_{\bf 0}$ denote the $-1/2$ subtraction.  This is the generator of the overall $U(1)_V$ symmetry. The other $Q_i$ generators  form closed commutation relations with combinations of $A_\bs + B_\bs$ given by
\be V_1 &=& -\frac{1}{2}(A_{(0,1,1)} + B_{(0,1,1)}) = \int d^3x\ \psi^\dagger\tau^1\psi \nn\\
V_2 &=& +\frac{1}{2}(A_{(1,0,1)} + B_{(1,0,1)}) = \int d^3x\ \psi^\dagger\tau^2\psi \nn\\
V_3 &=& -\frac{1}{2}(A_{(1,1,0)} + B_{(1,1,0)}) = \int d^3x\ \psi^\dagger\tau^3\psi \ .\label{vvvtaste}\ee
Combined, these have the commutation relations
\be  [V_i,V_j] = 2i\epsilon_{ijk} V_k \ ,\ \ \ [V_i,Q_j] = 2i\epsilon_{ijk} Q_k\ ,\ \ \ [Q_i,Q_j] = 2i\epsilon_{ijk}V_k\ .\ee
We can disentangle them by writing
\be J_L^i=\frac{1}{4}(V_i+Q_i)\ \ \ {\rm and}\ \ \ J_R^i = \frac{1}{4}(V_i-Q_i)\ .\ee
These are the generators of the $SU(2)_L\times SU(2)_R$ chiral symmetry, with $[J_L^i,J_L^j] = i\epsilon_{ijk}J_L^k$ and $[J_R^i,J_R^j] = i\epsilon_{ijk}J_R^k$ and $[J_L^i,J_R^j]=0$. 

\para
That leaves the axial  $U(1)_A$ unaccounted for. This was identified in \cite{aoki} as
\be Q_A = -\frac{1}{2}\left(A_{(1,1,1)} + B_{(1,1,1)}\right) = \int d^3x\ \psi^\dagger \Gamma \psi\ .\label{qa}\ee
Because this axial charge is constructed from the $A_\bs$ and $B_\bs$ generators, it is unquantised. Both the vector and axial $U(1)$ charges commute with the non-Abelian generators: $[Q_{\bf 0},J_L^i]= [Q_{\bf 0},J_R^i] = [Q_A,J_L^i] = [Q_A,J_R^i]=0$. The net result is that these generators form the $u(2) \oplus u(2)$ algebra seen in the continuum limit. This subgroup of the emanant symmetry was seen in \cite{aoki}. 

\para
However, there are further infrared symmetries. We still have the $Q_\bs$ from the second line of \eqn{3dtheta}, together with the combination $A_\bs - B_\bs$ with $\bs=(1,1,0)$, $(1,0,1)$, $(0,1,1)$ and $(1,1,1)$. Each of these mixes particles and anti-particles. For example, when written in terms of $\psi=\ft12(\alpha + i\beta)$, the charges $Q_\bs$ become
\be Q_{(0,1,1)} &=& \frac{i}{2}\int d^3x\ \left(\psi^T\tau^1\psi - \psi^\dagger \tau^1\psi^\star\right) \nn\\
Q_{(1,0,1)} &=& - \frac{i}{2}\int d^3x\ \left(\psi^T\tau^2\psi - \psi^\dagger \tau^2\psi^\star\right) \nn\\
Q_{(1,1,0)} &=&  \frac{i}{2}\int d^3x\ \left(\psi^T\tau^3\psi - \psi^\dagger \tau^3\psi^\star\right) \nn\\
Q_{(1,1,1)} &=&  \frac{i}{2}\int d^3x\ \left(\psi^T\Gamma\psi - \psi^\dagger \Gamma\psi^\star\right)
\ .\ee
The generators $A_\bs - B_\bs$ take the same form, but with the difference in the integrand replaced by a sum. 

\para
These additional generators do not commute with those of  $u(2) \oplus u(2)$. Instead, they form the group $SU(4)$, rotating four Weyl fermions of the same chirality. In addition, the $Q_A$ axial charge \eqn{qa} is central, but unquantised on the lattice. Indeed,  no-go theorems \cite{xf,ryan} state that it's not possible to construct a $Q_A$ that is both quantised and exponentially local.  It becomes periodic in the infrared, where all excitations carry quantised charge. 

\subsection*{Reduced Staggered Fermions}

There is an obvious reduction of our starting  Hamiltonian \eqn{3dabham}  in which we keep only the $a_\br$ Majorana operators and discard the $b_\br$. The low-energy theory is a single Dirac fermion. The lattice symmetries comprise only the unquantised $A_\bs$ operators defined in \eqn{3dAB}. In the infrared limit \eqn{3dabir}, only the anti-symmetric matrices $\Theta_\bs$ survive: there are four of them
\be A_{(0,1,1)} &\ra& -\frac{1}{2} \int d^3x\ \alpha^T\tau^1 \alpha\ ,\ \ \   \ \ A_{(1,0,1)} \ra +\frac{1}{2} \int d^3x\ \alpha^T\tau^2 \alpha\nn\\
A_{(1,1,0)} &\ra& -\frac{1}{2} \int d^3x\ \alpha^T\tau^3 \alpha\ ,\ \ \   \ \ A_{(1,1,1)} \ra -\frac{1}{2} \int d^3x\ \alpha^T\Gamma \alpha \ . \ee
The first three of these generate an $SU(2)$ symmetry and, because  $[\Gamma,\tau^i]=0$, the final generator provides a central $U(1)$. The upshot is that, again, we find the full $U(2)$ chiral symmetry. 

\para
The same low-energy content, namely two Weyl fermions, appeared in the Weyl semi-metal construction of Gioia and Thorngren \cite{ryan}. In that case, the precursor lattice generators were quantised, while here they are unquantised. Conversely, the staggered fermion has discrete rotational symmetry while the Weyl semi-metal does not.

\subsection{The Maximal Compact Subalgebra}\label{compactsec}

All of the generators of the Onsager algebra are ultra-local -- they are comprised of a sum of generators that have support only on a finite number of lattice sites. But they proliferate: if we start with the $U(1)$ particle number, generated by $Q_{\bf 0}$, then successive commutators with other generators result in the infinite-dimensional Onsager algebra. The largest ultra-local compact subgroup is simply $U(1)$. 

\para
But, as is usually the case with lattice fermions, there is a trade-off between locality and other properties. If we're willing to relax the locality condition, from ultra-local to merely exponentially local, then we can find a larger compact subgroup.

\para
To this end, first consider the following collection of charges, built from $A_\bs$ and $B_\bs$ defined in \eqn{3dAB}. 
\be {\cal K}  &=& \frac14\left( A_{(1,1,0)}+A_{(1,-1,0)}
+B_{(1,1,0)}+B_{(1,-1,0)} \right) \nn\\ &&\ \ \ +\ \frac12\left(
A_{(1,0,0)}+A_{(0,1,0)}
+B_{(1,0,0)}+B_{(0,1,0)} \right)\ .\ee
The operator ${\cal K}$ is part of the Onsager algebra, and hence conserved. We define the corresponding single-particle operator $K_{\br\br'}$ as
\be {\cal K} = i\sum_{\br,\br'} c_\br^\dagger K_{\br\br'} c_{\br'}\ .\ee
From this, we can construct something akin to the overlap operator
\be \Delta = \frac{K}{\sqrt{-K^2}}\; .\ee
One might worry that the denominator vanishes and this operator is ill-defined. Happily, that's not the case. In momentum space, $K$ takes the form
\be K(\bk) = i\left[\tau^3\cos(k_1a)\cos(k_2a) + \Gamma \tau^1\sin(k_1a) + \Gamma\tau^2\sin(k_2a)\right]\ee
where $\tau^i$ are the taste generators, defined in \eqn{taste}, and $\Gamma$ is given in \eqn{3dspin}.  Hence
\be K^2(\bk) = -\left(1+\sin^2(k_1a)\sin^2(k_2a)\right)\ee
and is nowhere vanishing. However, in constructing the operator $\Delta$ we  have sacrificed the ultra-locality of the Onsager algebra generators for a conserved charge that is only exponentially local. It obeys $\Delta^T=-\Delta$ and $\Delta^2=-\mathds{1}$.

\para
The pay-off is immediate however. We introduce the operators
\be J^+=\frac{1}{2} \sum_{\br,\br'} c^\dagger_\br \Delta_{\br\br'} c^\dagger_{\br'} \  , \ \   \ J^-= (J^+)^\dagger \ \ \ {\rm and} \ \ \ J^3 = \frac{1}{2}Q_{\bf 0}\ .\ee
These close among themselves, forming an $su(2)$ algebra
\be [J^3,J^\pm] = \pm J^\pm\ \ \ {\rm and}\ \ \  [J^+,J^-]=2J^3\ .\label{compactsu2}\ee
Moreover, because $\Delta^2=-1$, each of the Hermitian generators $J^1= \ft12 (J^++J^-)$, $J^2=-\ft{i}{2}(J^+-J^-)$ and $J^3$ is quantised. 
In addition, we can introduce a further $U(1)$ charge 
\be C =  i\sum_{\br,\br'} c_\br^\dagger \Delta_{\br\br'} c_{\br'}\ .\ee
This commutes with the $su(2)$ generators. This is a taste vector charge and is also quantised. (It's not the axial charge $Q_A$ defined in \eqn{qa}.)  In the infrared it reduces to the taste charge $-V_3$, defined in \eqn{vvvtaste}. 

\para
The upshot is that we have an exponentially local $u(2)$ compact subalgebra acting on the lattice fermions. This $u(2)$ subalgebra sits in the natural way inside $su(4)$; the four Weyl fermions that transform in the ${\bf 4}$ of $su(4)$ decompose  into the vector-like representation ${\bf 2}_{+1} \oplus {\bf 2}_{-1}$.

\subsection{Multiple Staggered Fermions}\label{multiplesec}

We can repeat this calculation with $N_f$ staggered fermions. Each site now has a collection of fermion operators $c_\br^p$ with $p=1,\ldots,N_f$. The Hamiltonian is
\be H = \frac{i}{2}\sum_{p=1}^{N_f}\sum_\br \sum_{i=1}^3\eta_i(\br) c_\br^{p\,\dagger}\left(c_{\br+\hat{\bf i}}^p - c_{\br - \hat{\bf i}}^p\right) \ .\ee
The Hamiltonian has an obvious $U(N_f)$ symmetry, rotating the different species. But this is part of  a larger Onsager algebra. We work in the Majorana basis, with $c_\br^p = \ft12 (a_\br^p + ib_\br^p)$. The conserved Onsager charges can now mix Majoranas of different species, so
\be Q_\bs^{pq} = \frac{i}{2} \sum_\br \xi_\bs(\br) a_\br^pb_{\br + \bs}^q\ee
with the signs $\xi_\bs(\br)$ given by \eqn{3dhorrid} as before. All of these commute with the Hamiltonian and all are integer quantised.  These close under commutation to an Onsager algebra that is the same as \eqn{3dons}, just dressed with flavour indices, where
\be A^{pq}_\bs &=& \frac{i}{2}\sum_\br \left[ \xi_\bs(\br) a^p_\br a^q_{\br +\bs} - \delta^{pq}\delta_{\bs,{\bf 0}}\right] \nn\\
B^{pq}_\bs &=& \frac{i}{2}\sum_\br \left[ \xi_\bs(\br) b^p_\br b^q_{\br +\bs} - \delta^{pq}\delta_{\bs,{\bf 0}}\right]  \ .\ee
In the infrared limit, we now get $8N_f^2$ charges 
\be Q_\bs^{pq} \ra \frac{i}{2} \int d^3x\ \alpha^{p\,T}\Theta_\bs \beta^q\ee
with the collection of 8 matrices $\Theta_\bs$ given in \eqn{3dtheta}. For the $A_\bs$ and $B_\bs$ charges, a symmetric $\Theta_\bs$ must be accompanied by an anti-symmetry in the $p,q$ indices, and vice-versa. The total number of generators -- $Q_\bs$, $A_\bs$ and $B_\bs$ -- is then $16N_f^2$. Among these is the  unquantised axial charge
\be Q_A = -\frac{1}{2} \sum_{p=1}^{N_f} \left(A_{(1,1,1)}^{pp} + B_{(1,1,1)}^{pp}\right)\ .\ee
The remaining infrared generators give the maximally chiral algebra $su(4N_f)$ acting on the purely left-handed Weyl fermions.

\para
We can also look at the maximal compact subalgebra. If we restrict to ultra-local generators, we have only the obvious $U(N_f)$, rotating the different flavours. Extending to exponentially local operators, as in Section \ref{compactsec}, this can be shown to become $U(2N_f)$. 

\section{Hamiltonian Wilson Fermions}\label{w3dsec}

In this section, we turn to Wilson fermions. We will again see that there is an exact lattice symmetry that emanates the full chiral symmetry in the infrared. This time, however, the simple Onsager algebra, derived from shifts of Majorana fermions, with its collection of ultra-local charges, is not sufficient: it provides only the vector symmetries of the theory. Instead, we must augment this algebra with further conserved charges. As we will see, there are two options to recover the chiral charges on the lattice: ultra-local, but unquantised; or quantised, but only exponentially local.

\para
As in the previous section, we work with a cubic lattice but, this time, each site houses a  collection of four fermion creation operators, $\psi^\dagger_{\br,\alpha}$, with $\alpha=1,2,3,4$ the spinor index, obeying the standard anti-commutation relations
\be \{\psi^\dagger_{\br,\alpha},\psi_{\br',\beta}\} = \delta_{\br,\br'}\delta_{\alpha,\beta}\ .\ee
We introduce gamma matrices $\{\gamma^\mu,\gamma^\nu\} = 2\eta^{\mu\nu}$ with $(\gamma^0)^\dagger = -\gamma^0$ and $(\gamma^i)^\dagger = \gamma^i$. Suppressing the spinor indices, the Wilson Hamiltonian  is
\be H= \frac{i}{2a}\sum_\br\sum_{i=1}^3\left[\psi^\dagger_\br\gamma^0\gamma^i \left(\psi_{\br+\hat{\bf i}} - \psi_{\br - \hat{\bf i}}\right) 
+\psi_\br^\dagger \gamma^0 \left(\psi_{\br + \hat{\bf i}} + \psi_{\br - \hat{\bf i}} + 2\left(\frac{ma}{3}-1\right)\psi_\br\right)\right]\,.\ \ \ \ \label{3dwilsonham}\ee
(Here we've picked the Wilson parameter to be, in the standard convention, $r=-1$; varying it  doesn't change the story below.) We've included the fermion mass $m$ in the Hamiltonian.

\para
In momentum space, we write
\be \psi_\alpha(\bk) = \frac{1}{\sqrt{N}} \sum_\br e^{-i\bk\cdot\br a} \psi_\br\ee
with $N$ the total number of lattice sites and where $\bk$ sits in the Brillouin zone
\be -\frac{\pi}{a} < k_i \leq \frac{\pi}{a}\  \ \ i=1,2,3.\ee
Note that each momentum spans twice the range of the reduced Brillouin zone \eqn{bz3d}. The Hamiltonian then takes the form
\be H = \int \frac{d^3k}{(2\pi)^3} \psi^\dagger(\bk)\, h(\bk)\, \psi(\bk)\ee
with
\be h(\bk) = -\frac{1}{a} \sum_{i=1}^3\gamma^0\gamma^i\sin(k_ia) + \frac{i}{a}\gamma^0\big(ma - W(\bk)\big)\label{w3dh}\ee
where $W(\bk)$ is the Wilson term which, in momentum space, reads
\be  W(\bk) = \sum_{i=1}^3(1-\cos(k_ia))\ \label{ww}.\ee
The energy spectrum is easily read off to be
\be E_\pm (\bk) = \pm\frac{1}{a} \sqrt{\sum_{i=1}^3\sin^2(k_ia) + \left(ma - W(\bk)\right)^2}\ .
\ee
For small momenta (and for $|ma|\ll 1$), the light fermion sits at the origin of the Brillouin zone, where we see the Hamiltonian for a single Dirac fermion emerging
\be h(\bk) = -\gamma^0\bgamma\cdot{\bk} + i m\gamma^0 + {\cal O}(k^2)\ .\ee
This becomes gapless when $m=0$. 

\subsection*{The Onsager Algebra}

The Onsager algebra again becomes apparent when  working with Majorana fermions. Suppressing the spinor index, we write
\be \psi_\br = \frac{1}{2} (a_\br + ib_\br)\ee
where $a_\br^\dagger = a_\br$ and $b_\br^\dagger = b_\br$ are now each four-component Majorana spinors. They satisfy
\be \{a_{\br,\alpha}, a_{\br',\beta}\} = \{b_{\br,\alpha},b_{{\br'},\beta}\} = 2\delta_{\br,\br'}\delta_{\alpha\beta}\ \ \ {\rm and}\ \ \ \{a_{\br,\alpha},b_{\br',\beta}\} = 0\ .\ee
We choose to work in a Majorana representation of the gamma matrices, so that all entries are real or, equivalently, $(\gamma^0)^T=-\gamma^0$ and $(\gamma^i)^T= \gamma^i$.  The Hamiltonian \eqn{3dwilsonham} then decomposes cleanly, with the $a_\br$ and $b_\br$ fermions decoupling
\be H &=& \frac{i}{4a}\sum_{\br,i} \left[a_\br^T\gamma^0(1+\gamma^i)a_{\br + \hat{\bf i}} + b_\br^T\gamma^0(1+\gamma^i)b_{\br + \hat{\bf i}}\right] 
\nn\\ &&\  \ \  +\  \frac{i(ma-3)}{4a}\sum_\br  \left[a_\br^T\gamma^0a_r + b_\br^T\gamma^0b_\br\right]\ .\ee
This is the property that ensures an Onsager algebra. Note that the decoupling between $a_\br$ and $b_\br$ occurs for any value of the mass $m$. This is a clear sign that the Onsager algebra has nothing to do with chiral symmetry in this model. 

\para
Again, $a_\br$ and $b_\br$ can be translated independently. (No annoying minus signs this time.) And again, we can use the obvious particle number symmetry $Q_{\bf 0}$ as a seed to generate a whole tower of charges. These are
\be Q_\bs  = \frac{i}{2}\sum_{\br} a_\br^Tb_{\br+\bs}\ . \ee
Commutators of these conserved charges generate new generators $A_\bs - B_\bs$ where
\be A_\bs  = \frac{i}{2} \sum_\br \left(a_\br^Ta_{\br+\bs} - 4\delta_{\bs,{\bf 0}}\right)\ \ \ {\rm and}\ \ \ B_\bs  = \frac{i}{2} \sum_\br \left( b_\br^T b_{\br + \bs} - 4\delta_{\bs,{\bf 0}}\right)\ .\label{wasbs}\ee
As in the previous section, the $Q_\bs$ are quantised while the $A_\bs$ and $B_\bs$ are not. The resulting Ons$_3$ algebra is then
\be [Q_\bs,Q_\bt] = -i(A_{\bs -\bt} - B_{\bs-\bt})\ \ \ {\rm and}\ \ \ [Q_\bs, A_\bt-B_\bt] =2 i\left(Q_{\bs-\bt} - Q_{\bs+\bt}\right)\label{wons1}\ee
together with $[A_\bs,A_\bt] = [B_\bs,B_\bt] = [A_\bs,B_\bt]=0$. 

\para
Despite the existence of the infinite-dimensional Onsager algebra on the lattice, not much becomes of it as we flow to the IR. There are now no matrices sitting within the charges to give us the extra structure, and we simply get
\be Q_\bs \ra \frac{i}{2}\int d^3x\ a^Tb= \int d^3x\ :\psi^\dagger \psi :\ .\ee
This generates the usual particle number $U(1)_V$ symmetry\footnote{This should be contrasted with \cite{ryan}, where the Weyl nodes sit at
$k_z=\pm K$ with $K\neq 0,\pi/a$, so that $b$-translations act differently on
the two nodes and generate an $SU(2)$ in the infrared. For Wilson fermions, the
single Dirac point sits at $\mathbf{k}=0$, where $\mathsf{T}_\mathbf{s}\to 1$,
so chirality must be supplied separately. We turn to this in Section~3.1.}. Meanwhile, the other generators amount to nothing at all: $A_\bs, B_\bs\ra 0$. 

\subsection{An Extended Onsager Algebra}

There is an additional symmetry on the lattice that becomes the chiral $U(1)_A$ in the infrared. The continuum chiral generator is $\gamma^5= i\gamma^0\gamma^1\gamma^2\gamma^3$, with $(\gamma^5)^\dagger=\gamma^5$, and fails to commute with the Hamiltonian \eqn{w3dh} because of the Wilson term
\be [\gamma^5,h(\bk)] = \frac{2i}{a}\gamma^0 (W(\bk)-ma)\gamma^5\ .\ee
Nonetheless, there is a simple fix. We can dress the chiral charge with momentum to construct a conserved charge. Here we follow the construction of Gioia and Thorngren \cite{ryan}. (A related construction was given in \cite{mit}.) We try the ansatz
\be Q_A = \int \frac{d^3k}{(2\pi)^3} \ \psi^\dagger(\bk) q_A(\bk) \psi(\bk)\ \ {\rm with}\ \ q_A(\bk) = \gamma^5\left(F(\bk) + i\sum_i \gamma^i v_i(\bk)\right)\,.\ \ \ \ \label{wchiral}\ee
with $F(\bk)$ and $v_i(\bk)$ real, periodic functions on the Brillouin zone. The charge is Hermitian. Requiring that it commutes with the Hamiltonian, $[h(\bk), q_A(\bk)]=0$, gives the condition
\be (W-ma) F = \sum_i \sin(k_ia) v_i\ .\ee
This requires that we set $m=0$, as one would expect. This is solved by 
\be F(\bk) = \frac{1}{8}\prod_i\left(1+\cos(k_ia)\right)\ \ \ {\rm and}\ \ \ v_i(\bk) = \frac{1}{8} \sin(k_ia) \prod_{j\neq i} \left(1+\cos(k_ja)\right)\ .\label{fandvi}\ee
The normalisation is chosen so that  $F(\bk={\bf 0})=1$ and $v_i(\bk={\bf 0}) = 0$, ensuring that $q_A(\bk)\approx \gamma^5$ near the origin of the Brillouin zone. 

\para
The chiral charge \eqn{wchiral} is not quantised, in agreement with the no-go theorems of \cite{xf,ryan}. It is, however, ultra-local. It can be written in position space as
\be Q_A = \sum_{\br,\br'} \psi_\br^\dagger (q_A)_{\br,\br'} \psi_{\br'}\label{wqalocal}\ee
where the only non-vanishing elements of $q_A$ are
\be (q_A)_{\br,\br + \brho}  =\frac{1}{2^{3 + |\brho|_1 } }\gamma^5 \left(1+ \sum_i \rho_i\gamma^i\right)\ \ \ {\rm with}\ \ \ \rho_i = 0,\pm 1\label{wqaposition}\ee
Here $|\brho|_1$ denotes the 1-norm, $|\brho|_1 = |\rho_1| + |\rho_2| + |\rho_3|$.

\subsection*{The Extended  Algebra}

The chiral charge $Q_A$ commutes with $U(1)_V$. This too, however, can be seen as part of a  larger Onsager algebra, extending \eqn{wons1}. To see this, note that in $q_A$ we have found a matrix (whether in momentum space or position space) that commutes with the Hamiltonian. That means that $q_A^n$ also commutes with the Hamiltonian, allowing us to construct an infinite tower of conserved charges
\be \tilde{Q}_n &=& \frac{i^{n+1}}{2} a^T q_A^nb\ .\ee
%
The factors of $i$ mean these are Hermitian and each commutes with the Hamiltonian.  Except for $\tilde{Q}_0$, which is particle number, none of them are quantised. (The difference with our previous construction is that shifts by the translation operator $\horridt_\bs$ implement a unitary transformation, while $b\ra iq_Ab$ does not.) 

\para
Moreover, because the Hamiltonian decomposes into $a_\br$ and $b_\br$ pieces, we can construct further conserved charges with $a^T q_A^na$ and $b^Tq_A^nb$. All of these can also be combined with the translations $\horridt_\bs$ used to construct the Onsager algebra \eqn{wons1}. We have
\be Q_{\bs,n} &=& \frac{i^{n+1}}{2}a^T\horridt_\bs q_A^n b \nn\\
A_{\bs,n} &=& \frac{i^{n+1}}{4}a^T\left(\horridt_\bs - (-1)^n \horridt_{-\bs}\right) q_A^n a\nn \\
B_{\bs,n} &=& \frac{i^{n+1}}{4}b^T\left(\horridt_\bs - (-1)^n \horridt_{-\bs}\right) q_A^n b\ .\label{wextend}\ee
These close among themselves to form the larger Ons$_{3,1}$ algebra
\be [Q_{\bs,n},Q_{\bt,m}] &=& -(-1)^m i (A_{\bs-\bt,n+m} - B_{\bs-\bt,n+m})\nn\\
{[}Q_{\bs,n},A_{\bt,m}] &=& i\left[(-1)^mQ_{\bs-\bt,n+m} - Q_{\bs+\bt,n+m}\right]\nn\\
{[}Q_{\bs,n},B_{\bt,m}] &=& -i\left[(-1)^mQ_{\bs-\bt,n+m} - Q_{\bs+\bt,n+m}\right]\ ,\label{wons31}\ee
together with $[A_{\bs,n},A_{\bt,m}] = [B_{\bs,n},B_{\bt,m}] = [A_{\bs,n},B_{\bt,m}] = 0$. The axial charge is given by 
\be Q_A = -\frac{1}{2}\left(A_{{\bf 0},1} + B_{{\bf 0},1}\right)\ .\ee
Now we can ask again: what happens in the infrared? This time, the low-energy limit has four distinct generators: these are
\be Q_{\bs,0} \ra \frac{i}{2}\int d^3x\ a^T b\ &,&\ \ \  Q_{\bs,1} \ra -\frac{1}{2}\int d^3x\ a^T\gamma^5b  \nn\\
A_{\bs,1} \ra -\frac{1}{2}\int d^3x\ a^T\gamma^5a\ &,&\ \ \ B_{\bs,1} \ra -\frac{1}{2}\int d^3x\ b^T\gamma^5b\ .\ee
The infrared generators $\ft 12Q_{{\bf 0},0}$, $\ft 12Q_{{\bf 0},1}$ and $\ft14(A_{{\bf 0},1}-B_{{\bf 0},1})$ form an $su(2)$ algebra: these are completed by the axial charge $Q_A = -\ft12 (A_{{\bf 0},1} + B_{{\bf 0},1})$ which generates $U(1)_A$, so that the extended Onsager algebra emanates the full $U(2)$ chiral symmetry of the massless Dirac fermion. 

\para
Here we've told the story for just a single Wilson fermion. With $N_f$ massless Wilson fermions, there is an obvious $U(N_f)$ symmetry. But the same structure that we saw in Section \ref{multiplesec} carries through: that $U(N_f)$ symmetry mixes with the Onsager algebra in an interesting way, and the end result is that the exact lattice symmetries emanate a $U(2N_f)$ chiral symmetry in the infrared.

\subsection*{What Price Locality?}

In Section \ref{compactsec}, we saw how we could exchange locality for a compact subalgebra. Here we will again see the same behaviour. But, in addition, there is a different compromise we can make, sacrificing locality of charges for quantisation.

\para
First the compact subalgebra. For this, we need the original Hamiltonian $h$, together with the axial charge generator $q_A$ defined in \eqn{wqaposition}. We combine them to form
\be \Delta = \frac{q_A+iah}{\sqrt{q_A^2 + a^2 h^2}}\ .\ee
Again, the denominator is nowhere vanishing. One can check that
\be  q_A^2+a^2h^2 = F^2+\sum_{i=1}^3 v_i^2  +\sum_{i=1}^3\sin^2(k_i a)+W^2\geq 1 \ ,
\ee
with $W$ the Wilson term \eqn{ww}, and the functions $F$, $v^i$ defined in \eqn{fandvi}. The operator $\Delta$ is therefore exponentially local, obeying $\Delta^T=-\Delta$ and $\Delta^\dagger\Delta = \mathds{1}$. We can construct conserved $su(2)$ charges
\be
 J^+=\frac12\sum_{\mathbf r,\mathbf r'}
 \psi^\dagger_{\mathbf r}\Delta_{\mathbf r\mathbf r'}
 (\psi^\dagger_{\mathbf r'})^T \ ,\ \ \  J^-=(J^+)^\dagger\ \ \ {\rm and}\ \ \ J^3=\frac12Q_{\bf 0}\ .\ee
 These obey the $su(2)$ commutation relations \eqn{compactsu2}. This time there is just a single Dirac fermion, or two Weyl fermions, in the infrared, and these transform in the ${\bf 2}$ of this  $su(2)$. The Witten anomaly prohibits gauging of this symmetry. In the present case, this is obstructed because, while $J^3$ is onsite, $J^\pm$ are only exponentially local. 

\para
If we take $N_f$ Wilson fermions, the construction above yields an $Sp(N_f)$ compact symmetry at finite lattice spacing. (As an aside: the $2N_f$ Weyl fermions transform in the fundamental representation of $Sp(N_f)$. If we take a subgroup $H\subset Sp(N_f)$, then this representation decomposes into a real representation of $H$, in the sense that any complex representation $R$ comes with its conjugate $\overline{R}$. In other words, this doesn't allow for chiral representations.)

\para
There is also a related construction that we can implement. The generators $J^\pm$ and $J^3$ above are all quantised. (Indeed, they must be if they are to form an $su(2)$ algebra.) But the quantisation of the Onsager charges  $Q_{\bs,n}$ for Wilson fermions was weaker than for staggered: they are all  ultra-local, but are quantised only for $n=0$. We can do better, and construct charges  $\hat{Q}_{\bs,n}$ that are quantised for all $\bs$ and $n$. The price we pay, as above, is that they are no longer ultra-local, merely exponentially local.

\para
To see this, we start with the axial charge $q_A$, either in momentum space \eqn{wchiral} or position space \eqn{wqaposition}. It is Hermitian, $q_A^\dagger = q_A$, and purely imaginary, $q_A^\star = -q_A$, and so  $q^T_A=-q_A$. We can then construct a new operator through the Cayley transform
\be U = \frac{1+iq_A}{1-iq_A}\ .\ee
This is real, $U^\star = U$, and orthogonal, $U^TU=\mathds{1}$. It commutes with the Hamiltonian $[h,U]=0$, and also with the $b$-translation operator $[\horridt_\bs,U]=0$. This operator is well-defined, which we can see by writing it as
\be U = \frac{1+ 2iq_A - q_A^2}{1 + q_A^2}\label{uq2}\ee
and noting that, in momentum space,  $q_A(\bk)^2 = F(\bk)^2 + \sum_i v_i(\bk)^2$  is positive semi-definite. Moreover, it's a suitable chirality operator since $U(\bk={\bf 0}) = i\gamma^5$.
The price we pay is that the operator $U$ has support on all lattice sites, but the support decays exponentially. To see this, note that the problem comes from the denominator in \eqn{uq2}, which can be expanded as
\be \frac{1}{1+q_A^2}  =\frac{1}{2} \sum_{n=0}^\infty \left(\frac{1-q_A^2}{2}\right)^n\ .\ee
As we've seen, $q_A$ is an operator that, in position space \eqn{wqaposition}, has range of at most one in each coordinate, so $q_A^{2n}$ has range of at most $2n$. But, because we have  $0\leq q_A(\bk) ^2\leq 1$, the position space operator has norm
\be \left|\left| \left(\frac{1-q_A^2}{2}\right)^n\right|\right| \leq \frac{1}{2^n}\ .\ee
This tells us that  all terms that have range $2n$ are suppressed by $1/2^n$. 

\para
We can use this new operator to define a new collection of conserved charges
\be
 \widehat Q_{\bs, n} &=& \frac{i}{2}a^T\horridt_\bs U^n b\nn\\
 \widehat{A}_{\bs,n} &=& \frac{i}{4} a^T\left(\horridt_\bs U^n - \horridt_{-\bs}U^{-n}\right) a\nn\\
 \widehat{B}_{\bs,n} &=& \frac{i}{4} b^T\left(\horridt_\bs U^n - \horridt_{-\bs} U^{-n}\right)b\ .\ee
Now, since $\horridt_\bs U^n$ is real and orthogonal, each of the $\widehat{Q}_{\bs,n}$ charges is quantised. The $\widehat{A}_{\bs,n}$ and $\widehat{B}_{\bs,n}$ generators are not. Once again, the axial charge $\widehat{Q}_A = -\ft12 (\whA_{{\bf 0},1} + \whB_{{\bf 0},1})$ is unquantised. The generators close under commutation to the $\rm{Ons}_4$ algebra
\be [\whQ_{\bs,n},\whQ_{\bt,m}] &=& - i (\whA_{\bs-\bt,n-m} - \whB_{\bs-\bt,n-m})\nn\\
{[}\whQ_{\bs,n},\whA_{\bt,m}] &=& i\left(\whQ_{\bs-\bt,n-m} - \whQ_{\bs+\bt,n+m}\right)\nn\\
{[}\whQ_{\bs,n},\whB_{\bt,m}] &=& -i\left(\whQ_{\bs-\bt,n-m} - \whQ_{\bs+\bt,n+m}\right)\ ,\ee
with all $\whA$ and $\whB$ commutators vanishing as usual. This algebra again emanates a $U(2)$ chiral symmetry.

\section{Hamiltonian Overlap Fermions}\label{over3dsec}

We now turn to overlap fermions and the associated Ginsparg-Wilson relation. Again, we work in the Hamiltonian formulation of the theory  \cite{creutz}. (See \cite{hersh} for a more recent discussion.)  

\para
We will again need the gamma matrices $\{\gamma^\mu,\gamma^\nu\} = 2\eta^{\mu\nu}$ with $(\gamma^0)^\dagger = -\gamma^0$ and $(\gamma^i)^\dagger = \gamma^i$ (so that we're working in mostly plus signature). 
We also take $\gamma^5= i\gamma^0\gamma^1\gamma^2\gamma^3$, which obeys $(\gamma^5)^\dagger = \gamma^5$ and $(\gamma^5)^2 = \mathds{1}$. 

\para
As in the previous section, we  introduce a 4-component complex fermion $\psi_\br$ at each lattice site. Later this will also carry a flavour index, but we stick to just a single fermion for now.  The overlap Hamiltonian takes the form
\be H   = \sum_{\br,\br'} \psi_\br^\dagger h_{\br\br'} \psi_{\br'}\ \ \ {\rm with} \ \ \ h=i\gamma^0D .\label{ham1}\ee
Here $D$ is the spatial Dirac operator. In the continuum, we would take $D=\gamma^i{\cal D}_i$, which clearly obeys $\{\gamma^5,D\}=0$. The key idea of overlap fermions is to relax this requirement, and choose instead a Dirac operator that obeys the Ginsparg-Wilson relation \cite{gw}
\be \{\gamma^5,D\} = aD\gamma^5 D\ \label{gw1}\ee
with $a$ the lattice spacing. It's further required that $D$ has the hermiticity properties
\be \gamma^5 D \gamma^5 = D^\dagger\ \ \ {\rm and}\ \ \ \gamma^0 D \gamma^0 = -D^\dagger\ .\ee
These relations imply  $D+D^\dagger=aD^\dagger D=aDD^\dagger$ and $h^\dagger=h$.

\para
We'll specify the form of $D$ below. (We'll specify it in momentum space, where its expression is simple; it is more complicated in position space as it is only exponentially local.) But already these algebraic conditions are sufficient to say something about the symmetries. 
The Hamiltonian has an obvious vector symmetry 
\be Q_V = \sum_{\br} \left(\psi^\dagger_\br \psi_\br - 2\right)\label{overqv},
\ee
but the beauty of the Ginsparg-Wilson relation is that there is also a chiral symmetry \cite{luscher}, defined by
\be Q_A = \sum_{\br,\br'} \psi_\br^\dagger q_{A\,\br \br'} \psi_{\br'}\ \ \ {\rm with}\ \ \ q_A= \gamma^5\left( \mathds{1}-\frac{a}{2}D\right)\ .\label{gwqa}\ee
This charge is conserved, with $[H,Q_A]=0$, but it's not quantised.  Indeed, one can check that
\be q_A^2 + \frac{a^2}{4}(\gamma^5 D)^2 = q_A^2 + \frac{a^2}{4} h^2 =  \mathds{1}\ .\label{gwcircle}\ee
At low momenta, $D\approx 0$ and $q_A$ approximates the continuum $\gamma^5$. But the eigenvalues of $q_A$ are momentum-dependent and, at the edges of the Brillouin zone, we have $q_A\approx 0$.  Trivially, this axial charge commutes with the vector charge: $[Q_V,Q_A]=0$.

\para
One can construct a different operator, usually called $\hat{\gamma}_5$, given by
\be \hat{\gamma}_5 = \gamma^5 (1-aD)\ .\label{5hat}\ee
The factor of two difference from $q_A$ is important. One can check that the Ginsparg-Wilson relation implies $\hat{\gamma}_5^2=\mathds{1}$, so this operator is quantised. It too commutes with the vector charge. It is not, however, conserved since $[h,\hat{\gamma}_5]\neq 0$. 

\para
Solutions to  the Ginsparg-Wilson relation can be written as \cite{niceg}
\be D = \frac{1}{a}\left( \mathds{1} + V \right)\ee
for some unitary operator $V$, $V^\dagger V  = \mathds{1}$, with hermiticity properties inherited from $D$, so that $\gamma^5 V \gamma^5 = V^\dagger$ and $\gamma^0 V \gamma^0 = -V^\dagger$.

\para
A solution to this in momentum space is the overlap operator
\be V(\bk) = \frac{X(\bk)}{\sqrt{X^\dagger(\bk)\,X(\bk)}}\ \ee
with $X(\bk)$ taken from the Wilson Hamiltonian \eqn{w3dh} (where we set the mass $ma=1$). Namely
\be X(\bk) = i\sum_{i=1}^3\gamma^i \sin(k_ia) + W(\bk)-1 \ \ {\rm and}\ \ W(\bk) = \sum_{i=1}^3\left(1-\cos(k_ia)\right)\ .\ee
The spectrum of the Hamiltonian is
\be E(\bk) = \pm\frac{1}{a}\sqrt{ 2\left( 1 + \frac{W(\bk)-1}{\sqrt{\sum_i\sin^2(k_ia) + (W(\bk) -1)^2}}\right)}\ .\ee
It vanishes only at the origin of the Brillouin zone, $\bk=0$. Here there is a single Dirac cone. When converted to position space, the Hamiltonian is only exponentially local \cite{hernandez}.

\subsection*{The Onsager Algebra and its Extension}

We split the Dirac spinor into real and imaginary pieces, as $\psi_\br = \frac{1}{2} (a_\br  + ib_\br)$. In the Majorana representation of gamma matrices, with  $(\gamma^0)^T=-\gamma^0$ and $(\gamma^i)^T= \gamma^i$, the position space Hamiltonian is real and once again decomposes into individual Hamiltonians for $a_\br$ and $b_\br$, and  translations of $b_\br$ alone are symmetries. As we've seen in previous sections, this immediately ensures that the following charges are symmetries of the Hamiltonian \cite{shuheng},
\be Q_\bs  = \frac{i}{2}\sum_{\br} a_\br^Tb_{\br+\bs} \ee
with $Q_{\bf 0}$ coinciding with the vector charge \eqn{overqv}.

\para
At this point, the story is almost identical to that in Section \ref{w3dsec}. There are further charges $A_\bs$ and $B_\bs$  defined in \eqn{wasbs} which, together with $Q_\bs$, form the Onsager algebra \eqn{wons1}. This Onsager algebra is not particularly exciting and reduces only to the trivial $U(1)_V$ symmetry in the infrared.

\para
However, as in Section \ref{w3dsec}, we can now include the chiral charge $q_A$ in the mix. This gives a larger collection of conserved charges, defined in \eqn{wextend}, which close to the Onsager algebra \eqn{wons31}. The only novelty is that the chiral charge $q_A$ is now taken to be the overlap operator \eqn{gwqa}. Everything in Section \ref{w3dsec} goes through in the same way, and this extended Onsager algebra emanates the full $U(2)$ chiral symmetry of a single gapless Dirac fermion in the infrared.

\para
Overlap fermions also naturally furnish the exact lattice subgroup we saw in the Wilson construction.
\para
To see this, define 
\be \Delta = q_A + \frac{ia}{2}h\ .\ee
Using the Majorana basis of  gamma matrices,  $q^T_A=-q_A$ and $h^T=-h$ and so $\Delta$ is also anti-symmetric
\be \Delta^T = -\Delta\ .\ee
Moreover, this is a unitary matrix by virtue of the condition \eqn{gwcircle}, together with $[q_A,h]=0$, so 
\be \Delta^\dagger \Delta = \mathds{1}\ .\ee
We then define the operators
 \be J^+ = \frac{1}{2} \sum_{\br,\br'}\psi_\br^\dagger \Delta_{\br\br'} (\psi^\dagger_{\br'})^T\ ,\ \ \ J^-=(J^+)^\dagger\ \ \ {\rm and}\ \ \ J^3=\frac{1}{2}Q_{\bf 0}\ee
It's simple to check that $J^\pm$ are conserved charges (it follows from $h\Delta + \Delta h^T=0$).  A direct calculation then shows that these obey the $su(2)$ commutation relations
\be [J^3,J^\pm] = \pm J^\pm \ \ \ {\rm and}\ \ \ [J^+,J^-] = 2J^3\ .\ee
Note that, while $J^3$ is purely onsite, the $J^\pm$ generators involve exponential tails. 

\para
If we have $N_f$ overlap fermions, there is an enlarged Onsager algebra, which emanates the full $U(2N_f)$ chiral symmetry in the infrared. Of this, an $Sp(N_f) \subset SU(2N_f)$ exists as a compact microscopic subgroup. As for Wilson fermions, the $J^\pm$ have exponential tails, which provides an obstruction to gauging, as expected given the Witten anomaly.

\section{Euclidean Staggered Fermions}\label{4dsec}

For high energy physicists, most simulations of relativistic quantum field theories take place with Euclidean spacetime lattices. In this section, we look at the Onsager symmetries, and their infrared limits, for Euclidean fermions. 

\para
There are, it turns out, exact Onsager symmetries for staggered, Wilson and overlap Euclidean fermions, each of which emanates the full chiral symmetry of the free fermions. The details differ somewhat from the Hamiltonian setting. Here we will discuss only Euclidean staggered fermions (see \cite{golterman} for a review) and exhibit the symmetries. The generalisation to Wilson and overlap fermions does not involve any ingredients beyond those contained in this section and the previous ones. 

\para
The first difference between spatial and spacetime lattices is the degrees of freedom themselves. We take a hypercubic lattice, with sites labelled by $\br  = (x_1,x_2,x_3,x_4) \in \Z^4$. We introduce a pair of Grassmann-valued degrees of freedom on each site, $\chi_\br$ and $\bar{\chi}_\br$. As usual in Euclidean space, these should be viewed as independent objects; the bar does not denote complex conjugation. They obey Grassmann anti-commutation relations
\be \{\chi_\br,\chi_{\br'}\} = \{\bar{\chi}_\br ,\bar{\chi}_{\br'}\} = \{\chi_\br,\bar{\chi}_{\br'}\} =0\ .\label{grassmann}\ee
These differ from the anti-commutation relations obeyed by the creation and annihilation operators $c_\br$ and $c_\br^\dagger$ given in \eqn{ccom}; there is no $\delta_{\br\br'}$ on the right-hand side of the final expression. 

\para
The Euclidean staggered fermion action is 
\be S = \sum_{\br,\br'} \bar{\chi}_\br D_{\br,\br'} \chi_{\br'}\label{4dact}\ee
with
\be  (D\chi)_\br = \sum_{\br'} D_{\br,\br'} \chi_{\br'} = \frac{1}{2a} \sum_{\mu=1}^4 \eta_\mu(\br) \left(\chi_{\br + \hat{\bmu}} - \chi_{\br-\hat{\bmu}}\right)\ee
and $\eta_\mu(\br)$ the usual collection of minus signs, now extended to four dimensions
\be \eta_1(\br) = -1 \ ,\ \ \eta_2(\br) = (-1)^{x_1}\ , \ \ \eta_3(\br) = (-1)^{x_1+x_2}\ ,\ \ \eta_4(\br) = (-1)^{x_1+x_2+x_3}\ .\ee
The finite-difference operator obeys $D^\star=D$ and $D^T=-D$, hence $D^\dagger = -D$. 

\para
To describe the infrared physics, we collect together the 16 fermions at the vertices of a unit hypercube by writing $\br = 2{\bf n} + \brho$ with $\rho_\mu = 0,1$ for $\mu = 1,2,3,4$.  We then work in the reduced Brillouin zone
\be -\frac{\pi}{2a} < k_\mu \leq \frac{\pi}{2a}\ee
The 16-component $\psi(\bk)$ is defined by putting together Grassmann operators from each corner and Fourier transforming. The components of $\psi$ are ordered so that, in analogy with \eqn{psi3d}, a Pauli matrix $X_\mu$, $Y_\mu$ or $Z_\mu$ with $\mu=1,2,3,4$ acts on the $\mu^{\rm th}$ bit of $\brho$. Correspondingly, the spinor variable $\psi$ obeys the patching condition across the Brillouin zone
\be \psi\left(\bk + \frac{\pi}{a}\hat{\bmu}\right)  = Z_\mu \psi(\bk)\ .\ee
In momentum space, the Dirac operator then becomes
\be D(\bk)  = \frac{i}{a} \sum_{\mu=1}^4 \gamma_\mu \sin (k_\mu a) \ ,\ee
where the $16\times 16$ dimensional matrices $\gamma^\mu$ follow the pattern of previous sections and are defined by
\be  \gamma_1 = -X_1\ ,\ \ \ \gamma_2 = Z_1X_2\ ,\ \ \ \gamma_3 = Z_1Z_2X_3\ ,\ \ \ \gamma_4 = Z_1Z_2Z_3X_4\ .\ee
The sole gapless mode sits at $\bk=0$. The 16-component fermion $\psi$ splits into four Dirac fermions -- the tastes --  with the $\gamma_\mu$, which obey $\{\gamma_\mu,\gamma_\nu\} = 2\delta_{\mu\nu}$, playing  the role of Euclidean gamma matrices. 

\para
What is the full chiral symmetry of Euclidean fermions in the continuum limit? Suppose that we have $N$ Euclidean, massless Dirac fermions: there is an obvious $U(N)\times U(N)$ chiral symmetry acting on the two chiralities. In Lorentzian signature, we can conjugate the right-handed fermions to turn them into left-handed fermions to reveal a larger $U(2N)$ symmetry, as we saw in Section \ref{3dsec}. But in Euclidean space, a right-handed fermion remains right-handed when conjugated, so it's not obvious that enlargement occurs. Nonetheless, the full chiral symmetry of $N$ Dirac fermions in Euclidean space is again $U(2N)$, but for a different reason: it hinges on the fact that  Dirac spinors $\Psi$ and $\bar{\Psi}$ should be treated as independent fields. 

\para
To see this, consider  $N$ Dirac spinors $\Psi_a$, with $a=1,\ldots,N$. (Here each $\Psi_a$ is the usual 4-component spinor.) We split these into left-handed and right-handed Weyl spinors that we write as $l_a$ and $r_a$
\be \Psi_a= \left(\begin{array}{c} l_a\\ r_a \end{array}\right)\  ,\ \ \bar{\Psi}_a = (\bar{l}_a,\bar{r}_a)\ .\ee
The continuum Euclidean action is
\be S = \int d^4x\ \sum_{a=1}^N \left(\bar{l}_a\sigma^\mu \partial_\mu r_a + \bar{r}_a \bar{\sigma}^\mu\partial_\mu l_a\right)\ee
with $\sigma^\mu = (\mathds{1},i\bsigma)$ and $\bar{\sigma}^\mu = (\mathds{1},-i\bsigma)$. 
The Euclidean rotation group is $Spin(4) \cong SU(2) \times SU(2)$. The Weyl fermions $l_a$ transform as $({\bf 2}, {\bf 1})$ and the $r_a$ as $({\bf 1},{\bf 2})$. Meanwhile, the Weyl fermions $\bar{l}_a$ transform as $(\bar{\bf 2},{\bf 1})$ and the $\bar{r}_a$ as $({\bf 1},\bar{\bf 2})$. (This differs from Lorentzian signature where complex conjugation flips the handedness.) But because the representation is pseudoreal, ${\bf 2}\cong \bar{\bf 2}$ and we can place  $l_a$ and $\bar{l}_a$  (and similarly $r_a$ and $\bar{r}_a$) on equal footing. Indeed, after an integration by parts, the action can be written as
\be S = \int d^4x\ \sum_{a=1}^{N} (\bar{r}_a,-r_a^T\epsilon) \bar{\sigma}^\mu \partial_\mu\left(\begin{array}{c} l_a \\ \epsilon \bar{l}_a^T\end{array}\right)\ee
where $\epsilon = i\sigma_2$ and we've used $\epsilon \bar{\sigma}^\mu\epsilon = -(\sigma^\mu)^T$. This form makes manifest the larger $U(2N)$ symmetry.

\subsection{The Onsager Algebra}

We will follow the path taken in the previous Hamiltonian models. As we will see, this will yield a closed Onsager algebra but, in contrast to the previous sections, it is not the full symmetry of the theory. We will then identify the further charges that enlarge the Onsager algebra. 
We define
\be \chi_\br = \frac{1}{2}(a_\br +ib_\br) \ \ \ {\rm and}\ \ \ \bar{\chi}_\br  =\frac{1}{2} (a_\br  - ib_\br)\ .\ee
These should be viewed as a simple change of integration variables. Since $\chi_\br$ and $\bar{\chi}_\br$ are viewed as independent, there is no reality condition on $a_\br$ and $b_\br$. As Grassmann variables, both square to $a_\br^2 = b_\br^2 = 0$,  in contrast to the Majorana operators in previous sections, which square to one. When written in terms of $a$ and $b$, the cross-terms vanish and we find that the action again decouples
\be S = \frac{1}{4}\sum_\br \left[a_\br (Da)_\br  + b_\br (Db)_\br\right]\ .\label{4dact2}\ee
This means that we can define a symmetry which translates  $b_\br$ by $\bs\in \Z^4$, while  leaving $a$ alone. We have
\be \horridt_\bs b_\br = \xi_\bs(\br) b_{\br + \bs} \ \  \ {\rm with}\ \ \ \xi_\bs(\br) = (-1)^{s_1(x_2+x_3+x_4)+s_2(x_3+x_4)+s_3x_4}\ .\ee
Note the difference in notation from \eqn{3dhorrid}. In previous sections, $\horridt_\bs$ was an operator on Hilbert space and so it acted on $b_\br$ (which was also an operator) by conjugation. But now $\horridt_\bs$ should be viewed as an infinite-dimensional matrix, acting on the vector of integration variables $b_\br$. 

\para
The action \eqn{4dact2} is also invariant under the symmetry that rotates $a_\br$ and $b_\br$. This acts in an onsite manner
\be Q_{\bf 0} a_\br  = b_\br \ \ \ {\rm and}\ \ \ Q_{\bf 0} b_\br = -a_\br\ .\ee
This is simply the $U(1)_V$ vector symmetry which acts in the usual way: $\chi_\br \ra e^{-i\alpha}\chi_\br $ and $\bar{\chi}_\br  \ra e^{i\alpha}\bar{\chi}_\br$.  The Onsager algebra is generated by these two, non-commuting symmetries: translations of $b_\br$, together with $U(1)_V$. We define the charges
\be Q_\bs  = \horridt_\bs Q_{\bf 0} \horridt_\bs^{-1}\ .\ee
These act as
\be Q_\bs  a_\br  = \xi_\bs(\br) b_{\br + \bs}\ \ \ {\rm and}\ \ \ Q_\bs b_\br  = - \xi_\bs(\br-\bs) a_{\br-\bs}. \ee
Note that $a_\br$ and $b_\br$ are translated in opposite directions. On $a_\br$, only the $\horridt_\bs$ acts while on $b_\br$ only the $\horridt_\bs^{-1}$ acts. Each of the operators $Q_\bs$ has eigenvalues $\pm i$. The factor of $i$ is because, in contrast to the charges defined in Section \ref{3dsec} and Appendix \ref{2dsec}, these charges  are anti-Hermitian, rather than Hermitian. The commutators of $Q_\bs$ close if we also include the transformations
\be A_\bs a_\br = \xi_\bs(\br) a_{\br + \bs} - \xi_\bs(\br-\bs) a_{\br-\bs} \ \ \ &{\rm and}&\ \ \ A_\bs b_\br = 0\nn\\
B_\bs b_\br = \xi_\bs(\br) b_{\br + \bs} - \xi_\bs(\br-\bs) b_{\br -\bs}\ \ \ &{\rm and}&\ \ \ B_\bs a_\br = 0\ .\label{4dab}\ee
Note that, in contrast to the Hamiltonian formalism, none of the transformations $Q_\bs$, $A_\bs$ or $B_\bs$ includes Grassmann bilinears. That's because, with anti-commutation relations \eqn{grassmann}, any two Grassmann bilinears commute. That's not true for fermionic creation and annihilation operators acting on a Hilbert space. The commutators of these transformations are 
\be [Q_\bs,Q_\bt] &=&\epsilon(\bt,\bt) \left[\epsilon(\bs,\bt) A_{\bs-\bt} -\epsilon(\bt,\bs)B_{\bs-\bt}\right] \nn\\
{[}Q_\bs,A_\bt{]} &=&-\epsilon(\bt,\bs) \left[\epsilon(\bt,\bt)Q_{\bs-\bt}-Q_{\bs+\bt}\right]\nn\\
{[}Q_\bs,B_\bt{]} &=&\epsilon(\bs,\bt) \left[\epsilon(\bt,\bt)Q_{\bs-\bt}-Q_{\bs+\bt}\right]\ \label{4dons}\\
{[}A_\bs,A_\bt{]} &=&\left[\epsilon(\bt,\bs)-\epsilon(\bs,\bt)\right] \left[A_{\bs+\bt}-\epsilon(\bt,\bt)A_{\bs-\bt}\right] \nn\\
{[}B_\bs,B_\bt{]} &=&\left[\epsilon(\bt,\bs)-\epsilon(\bs,\bt)\right] \left[B_{\bs+\bt}-\epsilon(\bt,\bt)B_{\bs-\bt}\right]\nn\ee
and $[A_\bs,B_\bt]=0$ where 
\be \epsilon(\bs,\bt) = (-1)^{s_2t_1+s_3(t_1+t_2)+s_4(t_1+t_2+t_3)}\ .\ee
There are $\pm i$ differences from the Hamiltonian Onsager \eqn{3dons} which can be traced to the fact that the operators $Q_\bs$, $A_\bs$ and $B_\bs$ are anti-Hermitian rather than Hermitian.

\subsubsection*{The Infrared Limit}

We write the 16-component fermion $\psi_\rho = \ft12(\alpha_\rho + i\beta_\rho)$ where $\rho$ is a 4-vector with components $\rho_\mu=0,1$, labelling the corners of the hypercube. In momentum space, the components $\alpha_\rho$ and $\beta_\rho$ are defined by
\be \alpha_\rho(\bk) &=& \frac{16}{\sqrt{N}} \sum_{\bf n} e^{-ia\bk(2{\bf n} + \brho)} a_{2{\bf n} + \brho}
\nn\\ \beta_\rho(\bk) &=& \frac{16}{\sqrt{N}} \sum_{\bf n} e^{-ia\bk(2{\bf n} + \brho)} b_{2{\bf n} + \brho}\ee
with $N$ the number of lattice sites. The charges $Q_\bs$ act on these fermions as
\be Q_\bs \alpha(\bk) = e^{ia\bk\cdot \bs} \Theta_\bs \beta(\bk) \ \ \ {\rm and}\ \ \ Q_\bs \beta(k) = -e^{-ia\bk\cdot\bs}  \Theta^T_\bs\alpha(\bk)\ee
with $\Theta_\bs$ the $16\times 16$ dimensional matrices defined as $\Theta_\bs = \Theta_1^{s_1} \Theta_2^{s_2}\Theta_3^{s_3}\Theta_4^{s_4}$, with
\be \Theta_1=X_1Z_2Z_3Z_4 \ ,\ \ \ \Theta_2 = X_2Z_3Z_4\ ,\ \ \ \Theta_3= X_3Z_4\ , \ \  \ \Theta_4= X_4\ .\label{4dtheta}\ee
These obey $\Theta_\bs \Theta_\bt = \epsilon(\bs,\bt)\Theta_{\bs+\bt}$.  Meanwhile, $A_\bs$ and $B_\bs$ defined in \eqn{4dab} act as
\be A_\bs \alpha(\bk) = \left(e^{ia\bk\cdot\bs} \Theta_\bs - e^{-ia\bk\cdot\bs}\Theta^T_\bs\right) \alpha(\bk)\ \ \ &{\rm and}&\ \ \ A_\bs \beta(\bk) = 0\nn\\
B_\bs \beta(\bk) = \left(e^{ia\bk\cdot\bs} \Theta_\bs - e^{-ia\bk\cdot\bs}\Theta^T_\bs\right) \beta(\bk)\ \ \ &{\rm and}&\ \ \ B_\bs \alpha(\bk) = 0\label{4dabab}\ee
In the continuum limit, the momentum-dependent phases become irrelevant and we have the local action
\be Q_\bs \alpha(\bk)  \ra \Theta_\bs \beta(\bk)\ \ \ {\rm and}\ \ \ Q_\bs \beta(\bk) \ra -\Theta^T_\bs \alpha(\bk)\ .\ee
Because we have $\Theta_\mu^2=\mathds{1}$ for each $\mu=1,2,3,4$, only the parity of the components $s_\mu$ is important. This means that we have 16 distinct charges $Q_\bs$. The action of the $A_\bs$ and $B_\bs$ matrices becomes
\be A_\bs \alpha(\bk) = (\Theta_\bs - \Theta_\bs^T)\alpha(\bk)\ \ \ &{\rm and}&\ \ \ A_\bs \beta(\bk) = 0\nn \\
B_\bs \beta(\bk) = (\Theta_\bs - \Theta_\bs^T)\beta(\bk)\ \ \ &{\rm and}&\ \ \ B_\bs \alpha(\bk) = 0\ .\ee
We see that the $A_\bs$ and $B_\bs$ survive the continuum limit only for those $\Theta_\bs$ that are anti-symmetric. Each $s_\mu= 0,1$ and one can check that $\Theta_\bs$ is anti-symmetric only for $\sum_\mu s_\mu = 2$ or $3$. There are 6 of the former and 4 of the latter, so there are 10 surviving $A_\bs$ and a further 10 surviving $B_\bs$. So in total, the infrared algebra closes on 36 generators. These generators span the Lie algebra $sp(4)$. (In the convention that $sp(1) \cong su(2)$.)

\para
The algebra $sp(n)$ can be defined as real anti-symmetric matrices acting on $\R^{4n}$ that commute with a quaternionic structure, meaning three anti-symmetric matrices ${\cal I}$, ${\cal J}$, and ${\cal K}$ that obey ${\cal I}^2 = {\cal J}^2 = {\cal K}^2 = -\mathds{1}$, together with ${\cal I}{\cal J} = - {\cal J}{\cal I} = {\cal K}$ and cyclic permutations. 

\para
Our generators, $Q_\bs$, $A_\bs$ and $B_\bs$, are best thought of as $32\times 32$ dimensional matrices, rather than $16\times 16$, since they act on  $(\alpha,\beta)^T$ as
\be Q_\bs  = \left(\begin{array}{cc} 0 & -\Theta_\bs \nn\\ \Theta_\bs^T & 0 \end{array}\right)\ \ \ {\rm and}\ \ \ A_\bs\pm B_\bs = \left(\begin{array}{cc}\Theta_\bs^T - \Theta_\bs & 0 \\ 0 & \pm (\Theta_\bs^T - \Theta_\bs)\end{array}\right)\ .\label{4d32}\ee
We will  see that they form a reducible representation of $Sp(4)$, with the appropriate $\gamma^5$ operator projecting onto the irreducible representations. First note that these commute with the spinor matrices
\be G_\mu = \left(\begin{array}{cc} \gamma_\mu & 0 \\ 0 & \gamma_\mu \end{array}\right)\ .\ee
We have $[\Theta_\bs,\gamma_\mu]=0$, and hence $[Q_\bs, G_\mu] = [A_\bs \pm B_\bs,G_\mu]=0$. We can construct three complex structures from these spinor matrices by taking ${\cal I} = G_1G_2$, ${\cal J} = G_1G_3$ and ${\cal K} = {\cal I}{\cal J}= -G_2G_3$. Each of these is anti-symmetric and squares to $-\mathds{1}$. This is what is required to form a quaternionic structure. By construction, each of them commutes with the generators $Q_\bs$, $A_\bs$ and $B_\bs$. This means that these generators do, as promised, provide a representation of $sp(4)$. 

\para
We can construct irreducible representations by introducing $G_5=-G_1G_2G_3G_4$, which obeys $G_5^T=G_5$, $G_5^2=\mathds{1}$ and $\{G_5,G_\mu\}=0$. Projecting with $\ft12 (\mathds{1} \pm G_5)$ then defines two subspaces $\R^{16}$ on which the generators act irreducibly.

\para
The continuum symmetry $Sp(4)$ is a subgroup of the full continuum chiral symmetry $U(8)$. But it is not the whole group. That differs from what we found for the Hamiltonian staggered fermion but, as we saw, something similar occurred for the Wilson and overlap fermions. As in those cases, we need an additional missing ingredient. 

\subsubsection*{The Enlarged Onsager Algebra}

The missing ingredient is well known: massless staggered fermions enjoy an additional exact $U(1)$ that we will refer to as  $U(1)_{\rm taste}$ \cite{karsten}. (It is often called $U(1)_\epsilon$ in the literature.) This  acts as 
\be \chi_\br \ra e^{i\alpha \varep(\br)}\chi_r\ \ \ {\rm and}\ \ \ \bar{\chi}_\br \ra e^{i\alpha \varep(\br)}\bar{\chi}_\br\ \ \rm{with} \ \ \varepsilon(\br) = (-1)^{x_1+x_2+x_3+x_4} \;.\ee
Acting on the $a_\br$ and $b_\br$ variables, this becomes
\be Q_{\rm taste} a_\br  = i\varepsilon(\br) a_\br\ \ \ {\rm and} \ \ \ Q_{\rm taste} b_\br = i\varepsilon(\br) b_\br\ .\label{4daxial}\ee
In the continuum limit this transforms two of the Dirac fermions with charge $+1$ and the other two with charge $-1$. (Its interpretation as a vector-like or axial symmetry is ambiguous: it depends on the mass term used to pair Weyl fermions \cite{vandoel,gs}.) It is not part of the Onsager algebra defined above, which makes us suspect that we're missing something. And we are.

\para
The clue comes from the presence of the alternating signs $\varep(\br)$ in the transformation above. We can dress other transformations with such signs to find new symmetries, effectively doubling our generators.  We start with 
\be \tilde{Q}_{\bf 0}a_\br = i\varep(\br) b_\br\ \ \ {\rm and}\ \ \ \tilde{Q}_{\bf 0} b_\br  = i\varep(\br)a_\br\ .\ee
Such a transformation isn't allowed in the Hamiltonian setting because it doesn't preserve the canonical anti-commutation relations, but it's fine when acting on Grassmann-valued integration variables $a_\br$ and $b_\br$. It is a symmetry of the action.  (It can be viewed as the axial symmetry, combined with a conjugation by $\horridt$ and a $\pi/4$ rotation by $Q_{\bf 0}$.) But given one new symmetry, we can construct a tower of symmetries in the usual way by conjugating with translations acting only on $b_\br$,
\be \tilde{Q}_\bs = \horridt_\bs \tilde{Q}_{\bf 0}\horridt_\bs^{-1}\ .\ee
These act as
\be \tilde{Q}_\bs a_\br = i\varep(\br) \xi_\bs(\br) b_{\br +\bs} \ \ \ {\rm and}\ \ \ \tilde{Q}_\bs b_\br = i\varep(\br-\bs)\xi_\bs(\br-\bs) a_{\br-\bs}\ .\ee
The factor of $i$ is there to ensure that, like $Q_\bs$, these are anti-Hermitian. Each of them obeys $\tilde{Q}_\bs^2 =-\mathds{1}$ and has eigenvalues $\pm i$.

\para
Commutators of the generators $\tilde{Q}_\bs$ close with $A_\bs$ and $B_\bs$, previously defined in \eqn{4dab}. This gives a second copy of the Onsager algebra
\be [\tilde{Q}_\bs,\tilde{Q}_\bt] &=&\epsilon(\bt,\bt) \left[\varep(\bs)\varep(\bt)\epsilon(\bs,\bt) A_{\bs-\bt} -\epsilon(\bt,\bs)B_{\bs-\bt}\right] \nn\\
{[}\tilde{Q}_\bs,A_\bt{]} &=&-\varep(\bt)\epsilon(\bt,\bs) \left[\epsilon(\bt,\bt)\tilde{Q}_{\bs-\bt}-\tilde{Q}_{\bs+\bt}\right]\nn\\
{[}\tilde{Q}_\bs,B_\bt{]} &=&\epsilon(\bs,\bt) \left[\epsilon(\bt,\bt)\tilde{Q}_{\bs-\bt}-\tilde{Q}_{\bs+\bt}\right]\ \ .\ee
However, the commutator of $Q_\bs$ with $\tilde{Q}_\bt$ gives us yet further symmetries
\be [Q_\bs,\tilde{Q}_\bt] = \varep(\bt)\epsilon(\bt,\bt) \left[-\varep(\bs) \epsilon(\bs,\bt)\tilde{A}_{\bs-\bt}+ \epsilon(\bt,\bs)\tilde{B}_{\bs-\bt}\right]\ee
where we've defined staggered (tilded) versions of $A_\bs$ and $B_\bs$ given by
\be \tilde{A}_\bs a_\br = i\left[\varep(\br) \xi_\bs(\br) a_{\br+\bs} + \varep(\br-\bs)\xi_\bs(\br-\bs) a_{\br-\bs}\right] \ \ \ &{\rm and}&\ \ \ \tilde{A}_\bs b_\br = 0 \nn\\
\tilde{B}_\bs b_\br = i\left[\varep(\br) \xi_\bs(\br) b_{\br+\bs} + \varep(\br-\bs)\xi_\bs(\br-\bs) b_{\br-\bs}\right] \ \ \ &{\rm and}&\ \ \ \tilde{B}_\bs a_\br = 0 \ .\ee
All further commutation relations among the set $Q_\bs$, $\tilde{Q}_\bs$, $A_\bs$, $\tilde{A}_\bs$, $B_\bs$ and $\tilde{B}_\bs$ close among themselves. We have
\be [Q_\bs,\tilde{A}_\bt] &=& \epsilon(\bt,\bs)\left[\varep(\bt)\epsilon(\bt,\bt)\tilde{Q}_{\bs-\bt} + \tilde{Q}_{\bs + \bt}\right]\nn\\
{[}Q_\bs,\tilde{B}_\bt] &=& -\varep(\bs)\epsilon(\bs,\bt)\left[\varep(\bt)\epsilon(\bt,\bt)\tilde{Q}_{\bs-\bt} + \tilde{Q}_{\bs + \bt}\right]\nn\\
{[}\tilde{Q}_\bs,\tilde{A}_\bt] &=& -\epsilon(\bt,\bs)\left[\epsilon(\bt,\bt){Q}_{\bs-\bt} + \varep(\bt){Q}_{\bs + \bt}\right]\nn\\
{[}\tilde{Q}_\bs,\tilde{B}_\bt] &=& \varep(\bs)\epsilon(\bs,\bt)\left[\varep(\bt)\epsilon(\bt,\bt){Q}_{\bs-\bt} + {Q}_{\bs + \bt}\right]\nn\\
{[}\tilde{Q}_\bs,{A}_\bt] &=& -\varep(\bt)\epsilon(\bt,\bs)\left[\epsilon(\bt,\bt)\tilde{Q}_{\bs-\bt} - \tilde{Q}_{\bs + \bt}\right]\nn\\
{[}\tilde{Q}_\bs,{B}_\bt] &=& \epsilon(\bs,\bt)\left[\epsilon(\bt,\bt)\tilde{Q}_{\bs-\bt} - \tilde{Q}_{\bs + \bt}\right]\ .\ee
Next, we have the commutators 
\be
{[}A_\bs,\tilde{A}_\bt{]} &=& (\epsilon(\bt,\bs) - \varep(\bs)\epsilon(\bs,\bt)) \left[\tilde{A}_{\bs + \bt} + \varep(\bt)\epsilon(\bt,\bt)\tilde{A}_{\bs-\bt}\right] \nn\\
{[}\tilde{A}_\bs,\tilde{A}_\bt{]} &=& (\varep(\bs)\epsilon(\bs,\bt) - \varep(\bt) \epsilon(\bt,\bs)) \left[{A}_{\bs+\bt} + \varep(\bt) \epsilon(\bt,\bt) {A}_{\bs-\bt} \right] \ .\ee
The commutators of $B$ and $\tilde{B}$ take the same form. Finally, we have $[A_\bs,B_\bt] = [A_\bs,\tilde{B}_\bt] = [\tilde{A}_\bs,B_\bt] = [\tilde{A}_\bs,\tilde{B}_\bt]=0$. 

\para
After all this, we can at last answer the question that motivated this enhancement: where is the taste  symmetry \eqn{4daxial}? The answer is that  it now sits within the extended Onsager algebra: it is  given by
\be Q_{\rm taste} = \frac{1}{2}\left(\tilde{A}_{\bf 0} + \tilde{B}_{\bf 0}\right)\ .\ee

\subsubsection*{The Enlarged Infrared Limit}

To take the infrared limit, we just need to chase down those minus signs in $\varep(\br)$. But this is straightforward: we have $\varep(2{\bf n} +\brho) = (-1)^{\rho_1+\rho_2+\rho_3+\rho_4}$. This is implemented on the 16-component spinors by the matrix $Z_1Z_2Z_3Z_4$. Ignoring the momentum-dependent phases, as usual, the low-energy action of the charges $\tilde{Q}_\bs$ is
\be \tilde{Q}_\bs \alpha \ra iZ_1Z_2Z_3Z_4\Theta_\bs \beta \ \ \ {\rm and}\ \  \ \tilde{Q}_\bs \beta \ra i\Theta_\bs^TZ_1Z_2Z_3Z_4\alpha\ ,\ee
with $\Theta_\bs$ defined in \eqn{4dtheta}. Similarly, $\tilde{A}_\bs$ and $\tilde{B}_\bs$ also pick up factors of $Z_1Z_2Z_3Z_4$. The result is that, acting on the 32-component spinors $(\alpha,\beta)^T$, as in \eqn{4d32}, we have the infrared action
\be \tilde{Q}_\bs= i\left(\begin{array}{cc} 0 & Z_1Z_2Z_3Z_4\Theta_\bs\\ \Theta_\bs^TZ_1Z_2Z_3Z_4 & 0 \end{array}\right)\ee
One can check that the infrared limit of the closed Onsager algebra generated by  $\tilde{Q}_\bs$, $A_\bs$ and $B_\bs$ gives another $Sp(4)$ symmetry at low energies. Here, however, we want to understand the infrared limit of the full, extended Onsager algebra. This includes the $\tilde{A}_\bs$ and $\tilde{B}_\bs$ generators, which become
\be  \tilde{A}_\bs \pm \tilde{B}_\bs = i\left(\begin{array}{cc} Z_1Z_2Z_3Z_4\Theta_s + \Theta^T_\bs Z_1Z_2Z_3Z_4 & 0 \\
0 & \pm(Z_1Z_2Z_3Z_4\Theta_\bs + \Theta_\bs^TZ_1Z_2Z_3Z_4)\end{array}\right)\ .\ \ \ \label{4dtildeab}\ee
There are 16 surviving $\tilde{Q}_\bs$ in the infrared. Meanwhile, the $\tilde{A}_\bs$ and $\tilde{B}_\bs$ survive only when the block-diagonal elements are symmetric: this holds when $\sum_{\mu=1}^4s_\mu = 0$, 3 or 4. This means that there are 6 surviving $\tilde{A}_\bs$ and six further $\tilde{B}_\bs$. (This is in contrast to the 10 $A_\bs$ and $B_\bs$.)

\para
The upshot is that we have 16 generators in $Q_\bs$, a further 16 in $\tilde{Q}_\bs$, 10 each in $A_\bs$ and $B_\bs$ and 6 each in $\tilde{A}_\bs$ and $\tilde{B}_\bs$. The infrared symmetry has 64 generators in total. In our conventions, each of these is anti-Hermitian and linearly independent. They therefore generate the full chiral symmetry $U(8)$ expected for four massless Euclidean Dirac fermions. 

\para
The central $U(1) \subset U(8)$ is generated, as in the Hamiltonian setting \cite{aoki}, by
\be Q_A = \frac{1}{2}\left(\tilde{A}_{(1,1,1,1)} + \tilde{B}_{(1,1,1,1)}\right) \ .\ee
This is unquantised. In the infrared limit, it coincides with $iG_5 = -iG_1G_2G_3G_4$, the product of gamma matrices acting on the 32-component spinors.

\para
As in previous sections, we can generalise the discussion to $N_f$ Euclidean staggered fermions, resulting in $4N_f$ gapless fermions in the continuum limit. The extended Onsager algebra now emanates a $U(8N_f)$ symmetry in the infrared. 

\para
Euclidean staggered fermions already admit an exact onsite $U(2N_f)$ symmetry on the lattice. (See, for example, \cite{hands}.) For $N_f=1$, its four generators are $Q_{\bf 0}$, $\tilde Q_{\bf 0}$, $\tilde A_{\bf 0}$ and $\tilde B_{\bf 0}$. There is also a {\it non-local} $U(4N_f)$ taste
symmetry of the free theory \cite{golterman}. This can be constructed
by taking the matrices $\Theta_\mu$ defined in \eqn{4dtheta} 
and dressing them with momentum-dependent signs,
\be
\widehat\Theta_\mu(\bk)
={\rm sign}[\cos(k_\mu a)]\,\Theta_\mu\ .
\ee
These obey the Clifford algebra $\{\widehat{\Theta}_\mu,\widehat{\Theta}_\nu\} = 2\delta_{\mu\nu}$ and have the correct periodicity across the reduced Brillouin zone. Their
products generate the full $u(4N_f)$ taste algebra. However, the discontinuities at the boundaries of the Brillouin zone mean that the generators are non-local, with algebraically decaying tails.

\subsection*{Reduced Staggered Fermions}

We can again look at reduced staggered fermions, with the action given by \eqn{4dact2} with the $b_\br$ integration variables discarded. This now describes two massless  Dirac fermions. In the context of lattice QCD, these were introduced in \cite{vandoel,gs}. The action enjoys the symmetries generated by $A_\bs$ and $\tilde{A}_\bs$ which, in the infrared, act as
\be A_\bs \alpha_\br \ra (\Theta_\bs - \Theta_\bs^T)\alpha_\br\ \ \ {\rm and}\ \ \ \tilde{A}_\bs \alpha_\br \ra i\left[Z_1Z_2Z_3Z_4\Theta_\bs  + \Theta_\bs^TZ_1Z_2Z_3Z_4\right] \alpha_\br \ .\ee
As described above, the action of the $A_\bs$ is non-trivial when $\Theta_\bs$ are anti-symmetric and those of $\tilde{A}_\bs$ is non-trivial when $Z_1Z_2Z_3Z_4\Theta_\bs$ are symmetric. There are ten of the former and six of the latter. 

\para
First the $A_\bs$. The surviving ten matrices are the products of any two of the following:  $\Theta_1$, $\Theta_2$, $\Theta_3$, $\Theta_4$, $\Theta_{(1,1,1,1)}$. These products form an $so(5) \cong sp(2)$ algebra. 

\para
Five of the surviving  $\tilde{A}_\bs$ are given by $\tilde{A}_{(0,0,0,0)}$, $\tilde{A}_{(1,1,1,0)}$, $\tilde{A}_{(1,0,1,1)}$, $\tilde{A}_{(1,1,0,1)}$, and $\tilde{A}_{(0,1,1,1)}$. These combine with the $A_\bs$ above to extend $so(5)$ to $so(6) \cong su(4)$. The final $\tilde{A}_\bs$ generator is $\tilde{A}_{(1,1,1,1)}$. This is central: it extends $su(4)$ to $u(4)$. Again, we see the full chiral symmetry emanating from the Onsager algebra. 

\section{Summary}
\label{summarysec}

For each of the fermionic formulations described above, the Onsager algebra emanates the full chiral symmetry in the infrared. But the details differ. Here we provide a summary.

\begin{itemize}
\item $N_f$ staggered fermions in the Hamiltonian formulation give rise to  $2N_f$ Dirac fermions in the infrared. For $d=3+1$ there is a magnetic ${\rm Ons}_3$ algebra at finite lattice spacing. All symmetry generators are ultra-local and a subset -- the $Q_{\bf s}$ -- are quantised.  In the infrared, this algebra emanates a $U(4N_f)$ chiral symmetry. 

The maximal  subalgebra of the Onsager algebra is the obvious $u(N_f)$, rotating the different flavours of fermions. But if we relax the ultra-local condition, to allow for exponentially local operators, we can construct a larger $U(2N_f)$ lattice symmetry.
\item $N_f$ Wilson fermions give rise to $N_f$ Dirac fermions in the infrared. There is an extended Onsager symmetry ${\rm Ons}_{3,1}$ at finite lattice spacing, which emanates the full $U(2N_f)$ chiral symmetry in the infrared.

The generators $Q_{\bs,0}$ are ultra-local and quantised. For other generators $Q_{\bs,n}$ with $n\neq 0$, we have a choice: we can either have them ultra-local and unquantised, or exponentially local and quantised. If we allow for exponentially local operators, we can also construct a $Sp(N_f)$ exact symmetry on the lattice.

\item $N_f$ overlap fermions in the Hamiltonian formulation give rise to $N_f$ Dirac fermions in the infrared.  As for Wilson fermions, there is an extended Onsager algebra ${\rm Ons}_{3,1}$ that emanates the full $U(2N_f)$ chiral symmetry. This time the Hamiltonian itself is only exponentially local, as is the  axial charge. These can be used to build  an exact $Sp(N_f)$ symmetry on the lattice.

\item
$N_f$ staggered fermions in the Euclidean formulation give rise to $4N_f$ Dirac fermions in the infrared. There is an extended Onsager algebra which emanates the full $U(8N_f)$ chiral symmetry. This time, there is nothing novel about the compact lattice symmetries: there is an onsite $U(2N_f)$, and non-local $U(4N_f)$.

\end{itemize}

Our discussion in this paper is limited to free fermions. It remains to understand what of this structure can survive different attempts at gauging.

\begin{appendices}

\section{Staggered Fermions in $d=2+1$ Dimensions}\label{2dsec}

In this appendix, we discuss the staggered fermion in $d=2+1$ dimensions. At low energies, the theory  describes two Dirac (i.e. two-component) fermions or, equivalently, four Majorana fermions. This model was discussed in the appendix of \cite{fermi}, where it was shown that it enjoys a magnetic Onsager algebra, with an emanant $SU(2)$ symmetry. Here we show that the lattice theory has a larger infrared symmetry, emanating a full $SO(4)$ symmetry rotating the Majoranas. 

\para
The construction, and much of the notation, follows that of Section \ref{3dsec}, so we give only an outline of the results. The Hamiltonian is
\be H = \frac{i}{2}\sum_\br \sum_{i=1}^2\eta_i(\br)c_\br^\dagger\left(c_{\br + \hat{\bf i}} - c_{\br - \hat{\bf i}}\right)\label{2dham}\ee
where $\hat{\bf i}$ denotes a shift by a single lattice site in the $i=1,2$ directions and $\eta_i(\br)$ are staggered signs, given by
\be \eta_1(\br) = -1\ \ \ {\rm and}\ \ \ \eta_2(\br) = (-1)^x\ .\ee%
This has the property that the product of signs around a unit plaquette is  $-1$.

\para
As in Section \ref{3dsec}, we package the fermions into a collection of four, one from each corner of a square. We do this by writing
%
%
\be \psi(\bk) = \big( c_{00}(\bk) ,  c_{01}(\bk) ,  c_{10}(\bk) ,  c_{11}(\bk)  \big)^T \ ,\label{psi2d}\ee
where the components are the  Fourier transforms of those fermions that sit on one corner of the square
\be c_{\brho}(\bk) = \sqrt{\frac{4}{N}} \sum_{\bf n} e^{-i\bk\cdot (2{\bf n}+\brho) a}c_{2{\bf n}+\brho}\ ,\ee
and $\brho$ is a two-component vector with components $\rho_i=0,1$. This means that we are again working in a reduced Brillouin zone, with
momenta lying in the range $-\pi/2a < k_i\leq \pi/2a$ for $i=1,2$. 

\para
The 4-component spinor $\psi$ will be acted on by a collection of Hermitian $4\times 4$ matrices which we build as tensor products of two Pauli matrices. As in Section \ref{3dsec}, there must be a patching condition as we cross the reduced Brillouin zone, with 
\be \psi\left(\bk  + \frac{\pi}{a}\hat{\bf i}\right) = Z_i \,\psi(\bk)\ .\label{2dperiod}\ee
The Hamiltonian in momentum space is given by
\be h(\bk) = -\sin (k_1a)\,X_1 + \sin (k_2a) \, Z_1 X_2 \ ,\label{hk2d}\ee
and, just as in Section \ref{3dsec}, there is a single gapless point in the reduced Brillouin zone, at $\bk = 0$. Here all four components of $\psi$ become gapless, meaning that we have two Dirac spinors. (In $d=2+1$, each Dirac spinor has two components.) At low energies, the Hamiltonian reduces to 
\be h(\bk) = a\bk\cdot\bgamma\ \ \ {\rm with}\ \ \gamma^1 = -X_1\ \ {\rm and}\ \ \gamma^2 = Z_1X_2\ee
These play the role of gamma matrices; they are accompanied by $\gamma^0 = iY_1X_2$ so that the collection obeys the usual Clifford algebra $\{\gamma^\mu,\gamma^\nu\} = 2\eta^{\mu\nu}$.

\para
The low-energy Hamiltonian enjoys a $U(2)$ symmetry, generated by the particle number symmetry $\psi \ra e^{i\alpha}\psi$, together with the $SU(2)$ generators
\be \tau^1 = X_1 Z_2\ ,\ \ \ \tau^2 = X_2\ ,\ \ \ \tau^3 = X_1Y_2\ .\label{2dtau}\ee
These obey $[\tau^i,\tau^j] = 2i\epsilon^{ijk}\tau^k$ and $[\tau^i,\gamma^\mu]=0$. This is the taste symmetry of the staggered fermions. 

\para
In fact, the continuum theory has a larger $SO(4) \cong [SU(2)\times SU(2)]/\Z_2$ symmetry. This is only apparent when the two Dirac fermions are written as four Majorana fermions. We will see the precursor of this symmetry sitting in the exact Onsager algebra on the lattice. 

\para
At first glance, it appears that all these are also  symmetries of the full lattice Hamiltonian \eqn{hk2d}; after all, they commute with the spinor matrices $\gamma^i$. But that's not the only requirement: the transformations must be consistent with the periodicity condition \eqn{2dperiod} and, because  $[\tau^i,Z^j]\neq 0$ none of the would-be $SU(2)$ symmetries are permissible when acting on the full Brillouin zone.

\subsection*{The Onsager Algebra}

When written in terms of Majorana fermions  $ c_\br = \ft12 (a_\br + i b_\br)$, the Hamiltonian again decouples into two pieces and is invariant under translations of $b_\br$ that leave $a_\br$ alone. The minus signs in the staggered fermion Hamiltonian \eqn{2dham} mean that we need compensating minus signs in the definition of the translation operator, which acts as
\be \horridt_1 b_\br\horridt_1^{-1} =  (-1)^y b_{\br + \hat{\bf 1}}  \ \ \ {\rm and}\ \ \  \horridt_2 b_\br \horridt_2^{-1} = b_{\br +\hat{\bf 2}} \ .\label{2dhorridt}\ee
We define translations that shift by a finite distance $\bs = (s_1,s_2)$ as $\horridt_\bs = \horridt_2^{s_2}\horridt_1^{s_1}$, so that  $\horridt_\bs b_\br \horridt_\bs^{-1} = (-1)^{s_1y}b_{\br + \bs}$.
We then use these to build the infinite tower of conserved, quantised charges
\be Q_\bs = \horridt_\bs Q_{\bf 0} \horridt_\bs^{-1} = \frac{i}{2}\sum_\br (-1)^{s_1y} a_\br b_{\br+\bs}\ .\label{qs2d}\ee
Commutators of these operators close with the introduction of two further operators
\be A_\bs = \frac{i}{2}\sum_\br \left[(-1)^{s_1y} a_\br a_{\br +\bs} - \delta_{\bs,{\bf 0}}\right] \ \ \ {\rm and}\ \ \ 
B_\bs = \frac{i}{2}\sum_\br \left[(-1)^{s_1y} b_\br b_{\br +\bs} - \delta_{\bs,{\bf 0}}\right]  \ .\ \ \ \label{2dabab}\ee
In contrast to the $Q_\bs$, the spectrum of the operators $A_\bs$ and $B_\bs$ is {\it not} quantised. 
The commutation relations are \cite{fermi}
\be [Q_\bs,Q_\bt] &=& -i\left[(-1)^{t_1(s_2+t_2)} A_{\bs-\bt} - (-1)^{t_2(s_1+t_1)} B_{\bs-\bt}\right]
\nn\\ {[}Q_\bs, A_\bt{]} &=& i(-1)^{s_1t_2}\left[(-1)^{t_1t_2}Q_{\bs-\bt} - Q_{\bs+\bt}\right]
\nn\\ {[}Q_\bs, B_\bt{]} &=& -i(-1)^{s_2t_1}\left[(-1)^{t_1t_2}Q_{\bs-\bt} - Q_{\bs+\bt}\right] \label{2dmagons}
\\ {[}A_\bs,A_\bt {]} &=& i\left((-1)^{s_2t_1} - (-1)^{s_1t_2}\right)\left[A_{\bs+\bt} - (-1)^{t_1t_2} A_{\bs-\bt}\right]
\nn \\ {[}B_\bs,B_\bt {]} &=& i\left((-1)^{s_2t_1} - (-1)^{s_1t_2}\right)\left[B_{\bs+\bt} - (-1)^{t_1t_2} B_{\bs-\bt}\right] \nn\ee
and $[A_\bs,B_\bt]=0$. 

\para
To take the infrared limit, we first convert the charges to momentum space. Writing  $\psi(\bk) = \ft12 \big(\alpha(\bk) + i\beta(\bk)\big)$, they become
\be Q_\bs =  \frac{i}{2} \int \frac{d^2k}{(2\pi)^2} \ \alpha(\bk)^\dagger e^{ia\bk\cdot \bs} (X_1Z_2)^{s_1}X_2^{s_2} \beta(\bk)\ .\label{2dqmom}\ee
Now we take the long wavelength limit. The tower  of charges $Q_\bs$ collapses to just four independent charges in the IR
\be Q_{(0,0)} \ra \frac{i}{2}\int d^2x\ \alpha^T \beta\ , \ \ \ &&Q_{(1,0)} \ra \frac{i}{2}\int d^2x\ \alpha^TX_1Z_2\beta\nn\\
Q_{(0,1)} \ra \frac{i}{2}\int d^2x\ \alpha^T X_2 \beta\ , \ \ \ &&Q_{(1,1)} \ra -\frac{1}{2}\int d^2x\ \alpha^TX_1Y_2\beta
.\ee
In addition, the IR closure requires that we keep 
\be A =  -\frac{1}{2}\int d^2x\ \alpha^T X_1Y_2 \alpha\ \ \ {\rm and} \ \ \ B = -\frac{1}{2}\int d^2x\ \beta^T X_1Y_2\beta \ .\label{2daaaa}\ee
These six generators close under commutation. This is simplest to see if we define 
\be J^1 &=& \frac{1}{2}Q_{(1,0)}\ ,\ \ \ J^2 = \frac{1}{2}Q_{(0,1)} \ ,\ \ \ J^3 =-\frac{1}{4}(A+B)\nn\\
 K^1 &=& \frac{1}{2}Q_{(0,0)}\ ,\ \ \ K^2 = -\frac{1}{2}Q_{(1,1)} \ ,\ \ \ K^3 =-\frac{1}{4}(A-B)\ .\ee
These form two commuting $su(2)$ algebras
\be [J^i,J^j] = i\epsilon^{ijk} J^k\ ,\ \ \ [K^i,K^j] = i\epsilon^{ijk} K^k \ ,\ \ \ [J^i,K^j]=0\ .\ee
This is the promised $so(4) \cong su(2) \oplus su(2)$ algebra. The $J^i$ generate the  $SU(2)$ taste symmetry \eqn{2dtau} that we saw previously: indeed, the generators $\tau^1$ and $\tau^2$ can be seen sitting in the definitions of $Q_{(1,0)}$ and $Q_{(0,1)}$. The second $su(2)$ is novel; when written in terms of $\psi$ and $\psi^\dagger$, it mixes particles and anti-particles. We can see this explicitly by replacing  $\alpha = \psi + \psi^{\star}$ and $\beta = -i(\psi - \psi^{\star})$ in the charges $Q_\bs$. They then become
\be Q_{(0,0)} \ra \int d^2x\ :\psi^\dagger \psi:\ , \ \ \ &&Q_{(1,0)} \ra \int d^2x\ \psi^\dagger X_1Z_2\psi\\
Q_{(0,1)} \ra \int d^2x\ \psi^\dagger X_2 \psi\ , \ \ \ &&Q_{(1,1)} \ra \frac{i}{2}\int d^2x\ \left[\psi^TX_1Y_2\psi - \psi^\dagger X_1Y_2\psi^\star\right]\nn
\ee
where the colons in the first term denote the subtraction of $1/2$  in \eqn{3dq0}. Clearly the final charge $Q_{(1,1)}$ is different here since it does not respect the complex structure of $\psi$. At an operational level, this arises simply because $X_1Y_2$ is anti-symmetric, while the other matrices are symmetric.

\subsubsection*{A Mass Deformation}

The continuum theory admits a mass deformation that preserves the full $SO(4)$ symmetry: one simply gives the same mass to all four Majorana fermions. (This mass necessarily breaks time reversal.)  It's natural to wonder if this same deformation preserves the full Onsager symmetry on the lattice. Here we show it does.

\para
The deformation of the Hamiltonian involves hopping across the diagonal
\be H_{\rm mass} &=& -\frac{im}{8}\sum_{\br} (-1)^x\left[a_\br (a_{\br +\hat{\bf 1} + \hat{\bf 2}} + a_{\br +\hat{\bf 1} - \hat{\bf 2}}) 
+ b_\br (b_{\br +\hat{\bf 1} + \hat{\bf 2}} + b_{\br +\hat{\bf 1} - \hat{\bf 2}})\right] \nn\\
&=& m\int \frac{d^2k}{(2\pi)^2} \ \cos(k_1a)\cos(k_2a) \psi^\dagger(\bk) \,Y_1X_2\,\psi(\bk)\ .\ee
From the first form, in position space, it's simple to check that the minus signs all cancel when implementing the translations of the $b_\br$ fermions \eqn{2dhorridt}. From the second form, in momentum space, we see that the energy spectrum becomes
\be E_\pm^2(\bk) = \sin^2(k_1a) + \sin^2(k_2a) + m^2 \cos^2(k_1a) \,\cos^2(k_2a)\ .\ee
This is gapped everywhere in the reduced Brillouin zone for non-zero $m$.

\section{Ginsparg, Wilson and Onsager}\label{gwosec}

Some years ago, Mandula pointed out that there is a different Onsager symmetry that one can associate to the Ginsparg-Wilson equation \cite{mandula0,mandula}. This differs from the algebra described in Section \ref{over3dsec}; in particular, it is {\it not} a symmetry of the Hamiltonian. In this appendix, we briefly describe this construction. 

\para
For reasons that will become apparent, we will rebrand the $\gamma^5$ matrix as $\Gamma$, so the Ginsparg-Wilson relation becomes
\be \{\Gamma,D\} = aD\Gamma D\ . \label{gw2}\ee
This has the usual solution
\be D = \frac{1}{a}\left( \mathds{1} + V \right),\ee
with $V$ obeying $V^\dagger V  = \mathds{1}$, $\Gamma V \Gamma = V^\dagger$ and $\gamma^0 V \gamma^0 = -V^\dagger$. This is sufficient to build the Onsager algebra \cite{mandula0,mandula}. Define
\be Q_n = \Gamma V^n\ \ \ {\rm and}\ \ \ G_n = -i\left(V^n- V^{\dagger\,n}\right)\ .\ee
These are Hermitian generators, with $Q_n^\dagger = Q_n$ and $G_n^\dagger = G_n$. Moreover, each of the $Q_n$ is quantised, with eigenvalues $\pm 1$, since $Q_n^2 = \mathds{1}$. Indeed,  $\hat{\gamma}_5$, defined in \eqn{5hat}, is given by $\hat{\gamma}_5=-Q_1$, while $q_A= \ft12 (Q_0-Q_1)$ which, because $[Q_0,Q_1]\neq 0$, is not quantised. 
A short calculation, using $Q_nQ_m = V^{m-n}$, shows that these operators form the ${\rm Ons}_1$ algebra
\be [Q_n, Q_m]= iG_{m-n}\ ,\ \ \ [G_n,G_m]=0\ ,\ \ \ [Q_n,G_m] = 2i(Q_{n-m}-Q_{n+m})\ .\ee
Note that we get the $1d$ Onsager algebra regardless of the dimension of the overlap fermions. It arises naturally from the Ginsparg-Wilson relation, simply through the existence of the unitary operator $V$.   

\para
In contrast to the examples considered in the main body of the paper, the overlap Hamiltonian does not commute with the Onsager algebra, although $h^2$ does.  One could search for a different Hamiltonian, one that commutes with the full Onsager algebra. There is  a natural candidate
\be h_{\rm ons} = \frac{i}{2}\Gamma\gamma^0(D- D^\dagger) =  \frac{i}{2a}\Gamma\gamma^0(V- V^\dagger) \ .\label{hons}\ee
This can also be written as 
\be h_{\rm ons} = hq_A \ .\ee
This new Hamiltonian indeed commutes with the full Onsager algebra $[h_{\rm ons},Q_n]=[h_{\rm ons},G_n]=0$. It's not the unique Hamiltonian that does this.

\para
There is a price to pay: the Hamiltonian  $h_{\rm ons}$ reintroduces doublers, something that the original Ginsparg-Wilson relation was designed to avoid. This is simple to see. With the Dirac operator written in the form $aD = \mathds{1} + V$, the original Hamiltonian $h_{GW} = i\gamma^0 D$ has zeros whenever $V$ has eigenvalue $-1$. But the new Hamiltonian $h_{\rm ons}$ has zeros whenever $V= \pm 1$. We can also see the doublers by writing
\be h_{\rm ons}^2 = h^2\left(\mathds{1} - \frac{a^2}{4}h^2\right)\ .\ee
The new Hamiltonian $h_{\rm ons}$ has zeros whenever $h=0$ or $h=\pm 2/a$. 

\para
We can see the connection between the Ginsparg-Wilson Hamiltonian $h$ and the Onsager-preserving Hamiltonian $h_{\rm ons}$ explicitly in $d=1+1$ dimensions. We take the gamma matrices $\gamma^0 = -i\sigma^1$, $\gamma^1 = -\sigma^3$ and $\Gamma = -\gamma^0\gamma^1 = -\sigma^2$. In momentum space, we start with the unitary operator
\be V = -\left(\begin{array}{cc} e^{ika} & 0 \\ 0 & e^{-ika} \end{array}\right)\ .\ee
The resulting overlap Hamiltonian is
\be h(k) = \frac{1}{a}\left[\sigma^1 \big(1-\cos(ka)\big) - \sigma^2 \sin(ka)\right]\ee
with spectrum $E_\pm(k) = \pm (2/a)\,\sin(ka/2)$. This has a single massless Dirac fermion at $k=0$. Meanwhile,
\be h_{\rm{ons}}(k) = \frac{1}{a}\sin{(ka)}\;\mathds{1},
\ee
which has two nodes at $k=0,\pi/a$.

\para
The connection between overlap fermions and Onsager-preserving Hamiltonians in higher dimensions appears to be less direct. Repeating the steps above, one finds an Onsager-preserving Hamiltonian that is not ultra-local: it includes exponentially decaying hoppings. 

\para
Relatedly, as we have seen, higher dimensional lattice models often exhibit higher dimensional Onsager algebras. In the language of this appendix, this means there are several commuting unitary operators, $V_iV_j = V_jV_i$. (For magnetic Onsager algebras that arise for staggered fermions, these unitaries are anti-commuting: $V_iV_j=-V_jV_i$.) It is not clear to us how one could use this in a Ginsparg-Wilson like construction.

\end{appendices}

\subsection*{Acknowledgements} 

We are grateful to Simon Catterall, Arkya Chatterjee, Sal Pace, Shu-Heng Shao, and Hersh Singh for helpful discussions. Both Claude and ChatGPT were used extensively to plough through the various commutation relations involved in the different algebras. They were significantly less helpful in writing the paper.
This work was supported by STFC grant ST/X000664/1 and a Simons Investigator Award.

\end{document}